\documentclass[fleqn,usenatbib]{mnras}

\usepackage{newtxtext,newtxmath}

\usepackage[T1]{fontenc}

\DeclareRobustCommand{\VAN}[3]{#2}
\let\VANthebibliography\thebibliography
\def\thebibliography{\DeclareRobustCommand{\VAN}[3]{##3}\VANthebibliography}

\usepackage{graphicx}	
\usepackage{amsmath}	
\usepackage[utf8]{inputenc}

\title[Stellar density benchmark stars]{Benchmark stars for mean stellar density and surface gravity estimates of solar-type stars}

\author[P. F. L. Maxted]{
P. F. L. Maxted$^{\href{https://orcid.org/0000-0003-3794-1317}{\includegraphics[scale=0.5]{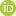}}}$\thanks{E-mail: p.maxted@keele.ac.uk}\\
Astrophysics Group, Keele University, Staffordshire ST5 5BG, UK\\
}
\date{Accepted XXX. Received YYY; in original form ZZZ}

\pubyear{2024}

\begin{document}
\label{firstpage}
\pagerange{\pageref{firstpage}--\pageref{lastpage}}
\maketitle

\begin{abstract}
Adding an independent estimate of the mean stellar density, $\rho_{\star}$, as a constraint in the analysis of stars that host transiting exoplanets can significantly improve the precision of the planet radius estimate in cases where the light curve is too noisy to yield an accurate value of the transit impact parameter, e.g. the light curves of Earth-size planets orbiting in the habitable zone of Sun-like stars that will be obtained by the PLATO mission.
I have compiled a sample of 36 solar-type stars for which analysis of high-quality light curves together with constraints on the orbital  eccentricity yield mean stellar density measurements with a median error of 2.3\,per~cent. Of these, 8 are in transiting exoplanet systems and 28 in eclipsing binary systems with very low mass companions that contribute $<0.1$\,per~cent of the total flux in the V band. A re-calibrated empirical relation for stellar mass as a function of T$_{\rm eff}$, $\rho_{\star}$ and [Fe/H] has been used to find mass estimates with a typical precision of 5.2\,per~cent for the stars in this sample.
Examples are given of how this sample can be used to test the accuracy and precision of $\rho_{\star}$ and $\log g$ estimates from catalogues of stellar parameters for solar-type stars.
\end{abstract}




\section{Introduction}
\label{sec:intro}

For any star in a binary system, we can use Kepler's 3$^{\rm rd}$ law to show that 
\begin{equation}
\label{eqn:rhostar}
  \mbox{$\rho_{\star}$} =
   \frac{3\mbox{M$_{\star}$}}{4\pi R_{\star}^3} =
   \frac{3\pi}{GP^2(1+q)}
    \left(\frac{a}{\mbox{R$_{\star}$}}\right)^3,
  \end{equation}
 where $P$ and $a$ are the period and semi-major axis of the Keplerian orbit,
and $q = M_2/\mbox{$M_{\star}$}$ is the mass ratio for a companion
with mass $M_2$ to a star with mass $M_{\star}$ and radius
$R_{\star}$.
The observation that the mean stellar density of the stars in an eclipsing binary system can be determined from an analysis of the light curve given an estimate of the mass ratio, $q$, goes back at least as far as \citet{1899ApJ....10..308R} and \citet{1899ApJ....10..315R}.

For transiting exoplanet systems with a transit impact parameter $b$, we can use the approximation $q\rightarrow 0$ in equation (\ref{eqn:rhostar}) and assume that the star and planet are spherical bodies with radii $R_{\star}$ and $r_{\rm pl} = k\,R_{\star}$, respectively,  to derive the following relation between the transit duration, $T_{14}$, and the mean stellar density that applies in the case that the orbit is circular \citep{2003ApJ...585.1038S}:
\begin{equation}
\tilde{\rho} = \left(\displaystyle \frac{4{\pi }^{2}}{{P}^{2}G}\right){\left(\displaystyle \frac{{\left(1+k\right)}^{2}-{b}^{2}\left(1-{\sin }^{2}[\pi {T}_{14}/P]\right)}{{\sin }^{2}[\pi {T}_{14}/P]}\right)}^{3/2}.
\label{eqn:rhostar0}
\end{equation}
For the more general case of a planet on an eccentric orbit with eccentricity $e$ and longitude of periastron\footnote{Care is needed to distinguish whether the value of $\omega$ given in published study applies to the star or the planet since these differ by 180$^{\circ}$.} $\omega$, the transit impact parameter is 
\[b=\displaystyle \frac{a\cos i}{{R}_{\star }}\left(\displaystyle \frac{1-{e}^{2}}{1+e\sin \omega }\right).\]
There is no exact version of equation (\ref{eqn:rhostar0})  for eccentric orbits, but the following approximation is valid for small values of $e$ \citep{2014MNRAS.440.2164K}:
\begin{equation}
\rho_{\star} \approx \frac{(1-e^2)^{3/2}}{(1+e\sin \omega )^{3}} \times\tilde{\rho} = \psi\times\tilde{\rho}.
\label{eqn:rhotilde}
\end{equation}

The existence of a relation between transit duration and mean stellar density is very useful for the analysis of transiting exoplanet light curves. For the case of transits with low signal to noise, including an independent estimate of $\rho_{\star}$ as a prior in the analysis helps to reduce correlations in the posterior probability distributions between $R_{\star}/a$ and $b$, and so improves the accuracy and precision of the measured value of $k$ \citep[e.g. ][etc.]{2013ApJ...767..127H, 2022MNRAS.514...77M}.  
This will be particularly relevant for PLATO mission  \citep[PLAnetary Transits and Oscillations of stars, ][]{2024arXiv240605447R}, which will find many small planets orbiting bright stars where $\rho_{\star}$\ can be estimated using asteroseismology (P1 sample). 
PLATO will also observe many faint stars (P5 sample) for which asteroseismology will not be possible. 
The mean stellar density for these stars will be estimated from analysis of the stellar spectrum, catalogue photometry and the parallax.  
Some care is needed to account for the possibility of an eccentric orbit if the analysis of the light curve is done assuming a circular orbit \citep{2022AJ....164...92G}.
For transits observed at high signal-to-noise with good constraints on the orbital eccentricity, the mean stellar density inferred from the light curve analysis can be combined with stellar models and other measurements to infer the stellar mass and, hence, the radius of the star and planet to good precision \citep{2023AJ....166..132E, 2015A&A...575A..36M}. 
It is also possible to use empirical relations to estimate $M_{\star}$ from $\rho_{\star}$ \citep{2010A&A...516A..33E}.

In the ``DEV'' (development) phase of PLATO prior to launch, a set of stellar density benchmark stars will be useful for testing the models and algorithms used to estimate $\rho_{\star}$ from spectroscopy, photometry and the parallax. 
This is clearly relevant to the P5 sample, but also for the P1 sample because this estimate of $\rho_{\star}$ will be used to set a prior for the interpretation for the power spectrum. 
During the ``OPS'' (operational) phase, stellar density benchmarks in the PLATO field of view that are also in the P1 sample can be used to validate the  estimates of $\rho_{\star}$ from asteroseismology. 
PLATO aims to measure planet radii to better than 3\%, so $\rho_{\star}$ needs to be accurate to a few percent to avoid a significant contribution to the error budget.

The dependence on the orbital eccentricity shown in equation (\ref{eqn:rhotilde}) is quite strong so $e$ needs to be known to an accuracy of $\lesssim$\,0.005 to achieve a precision of a few percent on $\rho_{\star}$. 
This is possible for hot-Jupiter systems with good radial velocity measurements where the phase of the secondary eclipse has been detected in the light curve. 
The offset from orbital phase 0.5 relative to the time of mid-transit of the secondary eclipse after correction for the light travel time across the orbit is approximately $e\cos{\omega}$ for near-circular orbits. 

Both equations  (\ref{eqn:rhostar0}) and  (\ref{eqn:rhotilde}) use the approximation $q\rightarrow0$. 
For companions where this approximation is not valid (brown dwarfs and stars), the same equations can be used provided that a factor $(1+q)$ seen in equation (\ref{eqn:rhostar}) is included in the denominator.
It is now possible using \'{e}chelle spectrographs on large telescopes operating at near-infrared wavelengths to measure $q$ directly in eclipsing binary (``EB'') systems composed of a solar-type star and a low-mass (``LM'') M-dwarf companion (``EBLM systems'') where the flux ratio in the optical band is $\lesssim 0.2$\,per~cent. 
These are ideal benchmark stars because the mass and radius for both stars and the effective temperature of the primary star, $T_{\rm eff,1}$, can be measured to good precision and accuracy with negligible dependency on stellar models. 
The flux contribution from the M-dwarf is small enough to be ignored for the analysis of observations at optical wavelengths for these EBLM systems. 
The properties of the low-mass companion in these EBLM systems can be estimated accurately enough for the apparent magnitudes of the solar-type star to be calculated with negligible additional uncertainty after correcting for the small flux contribution from the M-dwarf.

EBLM systems where the spectroscopic orbit is only available for primary star (SB1 systems) can also be useful as stellar density benchmark stars because $q$ can be estimated from the mass function given any reasonable estimate for the primary star mass, $M_{\star}$. 
For a typical EBLM systems with  $q\approx 0.2$, an error of 10\% in the estimate of  $M_{\star}$ implies an error $\lesssim 1$\% in $\rho_{\star}$. 
To estimate $q$, I have used an updated version of the empirical relation $M_{\star}(\rho_{\star}, {\rm T}_{\rm eff}, {\rm [Fe/H]})$ from \citet{2010A&A...516A..33E}. 
This process can be iterated to obtain an estimate of $q$ that is consistent with the values of $M_{\star}$ and $\rho_{\star}$. 
This method has the advantage that it does not depend on stellar models and we can account for the observed scatter $\approx 5$\% around the assumed $M_{\star}(\rho_{\star}, {\rm T}_{\rm eff}, {\rm [Fe/H]})$ relation. 

The recalibration of the empirical relation $M_{\star}(\rho_{\star}, {\rm T}_{\rm eff}, {\rm [Fe/H]})$ is described in Section~\ref{sec:recalib}. The selection of benchmark stars and, where necessary, re-analysis of the available observations to derive accurate values of $\rho_{\star}$ is described in Section~\ref{sec:sample}. Discussion of the results and examples of using this sample to test the accuracy of $\rho_{\star}$ and $\log g$ estimates obtained from published catalogues can be found in Section~\ref{sec:discuss}. Conclusions are given in Section~\ref{sec:conclusions}.
All values of the mean stellar density are quoted relative to the Sun in nominal solar units \citep{2016AJ....152...41P}, i.e. $\rho_{\odot} = 1409.826$\,kg\,m$^{-3}$.

\begin{figure}
    \centering
    \includegraphics[width=1\linewidth]{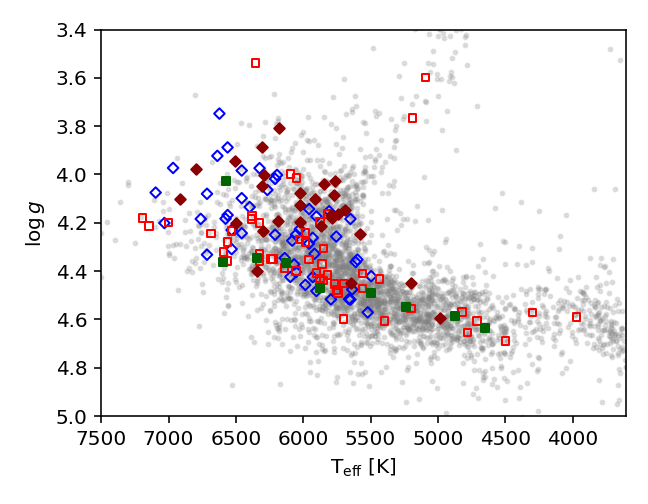}
    \caption{Distribution of stars selected from DEBCat to re-calibrate the empirical mass and radius polynomial functions. The primary and secondary stars in each binary are shown as open diamond (blue) and open square symbols (red), respectively. Planet host star parameters from SWEET-Cat are plotted as small points (grey) in the image background. EBLM systems and planet host stars from Table~\ref{tab:summary} are plotted as filled diamonds (red) and squares (green), respectively.}
    \label{fig:Teff_logg_DEBCat}
\end{figure}

\section{Re-calibration of empirical mass and radius polynomial functions}
\label{sec:recalib}

\citet{2010A&ARv..18...67T} derived empirical relations for the mass and radius of stars expressed as polynomial functions of their effective temperature, surface gravity and metallicity based on their own compilation of mass and radius measurements accurate to 3\,per~cent or better for stars in binary systems. 
\citet{2010A&A...516A..33E} derived an empirical relation for the mass and radius of stars as polynomial functions of their mean stellar density, effective temperature and metallicity based on the same data.
There are now over 350 eclipsing binary star systems with mass and radius measurements of comparable quality listed on the DEBCat website \citep{2015ASPC..496..164S}.\footnote{\url{https://www.astro.keele.ac.uk/jkt/debcat/}}
The quantity and quality of measurements for late-type stars in long-period eclipsing binary systems has improved substantially compared to the data available to \citeauthor{2010A&ARv..18...67T} and \citeauthor{2010A&A...516A..33E} so I have used the data from DEBCat downloaded on 2025-03-05 to derive new empirical relations that are suitable for estimating the mass of a typical planet host star, i.e. a slowly-rotating, late-type main-sequence or subgiant star. 
I removed systems containing pre-main-sequence stars and stars outside the effective temperature range $3940\,{\rm K} <  {\rm T}_{\rm eff} < 7220\,{\rm K}$. This effective temperature range corresponds to a spectral type range F0\,V to K9\,V. This covers the full range of ${\rm T}_{\rm eff}$ for the benchmark stars discussed below and the majority of known planet host stars.  
I also removed metal-poor stars from the sample ([Fe/H] $<-1$)  because the empirical relations defined by \citeauthor{2010A&ARv..18...67T} and \citeauthor{2010A&A...516A..33E} include only a linear term in [Fe/H] and the primary use for these empirical relations will be to estimate the properties of stars with solar-like metallicity. 
I computed the oblateness of the stars in each binary system using equation (43) from \citet{1933MNRAS..93..462C} and excluded stars from the sample with an oblateness $>0.0025$ and all stars in systems with orbital periods $P<5$\,days. 
This excludes stars that are significantly distorted by the gravity of their companions and stars that rotate rapidly because the tidal interaction between the stars is strong enough to force them to rotate synchronously with the orbit. 

For the dwarf components of the eclipsing red-giant + dwarf-star systems  KIC~8410637 and KIC~5640750 I use the values of $T_{\rm eff,2}$ from Table~4 of \citet{2018MNRAS.478.4669T} rather than the values quoted in DEBCat. 
These latter values were taken from Table~3 of \citeauthor{2018MNRAS.478.4669T} but are not accurate because the central surface brightness ratio was used instead of the average surface brightness ratio to estimate $T_{\rm eff,2}$ (Southworth, priv. comm.).\footnote{The central surface brightness ratio is not the same as the average surface brightness ratio for stars with different limb-darkening profiles.}
For KIC~6131659, I use the masses and radii derived from my own analysis of the {\it Kepler} light curve described in Appendix~\ref{sec:kic}.
The binary systems EBLM~J0113+31 and EBLM~J0608$-$59 have been excluded from the calibration data because these two systems have directly-measured mass ratios and so can be used as a test of the reliability of the method described below to estimate stellar mass and mean stellar density for the solar-type star in EBLM systems.

The calibration data for 98 stars in 58 eclipsing binary systems used are given in Table~\ref{tab:debcat}. 
The distribution on these stars in the Kiel diagram (T$_{\rm eff}$ -- $\log g$  plane) is shown in Fig.~\ref{fig:Teff_logg_DEBCat}.
An unweighted least-squares fit to these data for the same polynomial functions as used by \citet{2010A&ARv..18...67T} and \citet{2010A&A...516A..33E} results in the following empirical relations that can be used to estimate the mass and radius of FGK dwarf and subgiant stars using the variables  $X = \log_{10}({\rm T}_{\rm eff}/K) -4.1$, $Y = \log_{10}(\rho_{\star}/\rho_{\odot})$, and $\log g$ in cgs units.
\begin{equation}
\label{eqn:TorresM}
\begin{split}
\log_{10}(M/M_{\odot}) = & ~2.424296 + 2.14856\,X -0.525793\,X^2 \\
& -2.602827\,X^3 -0.257692\,(\log g)^2  \\
& + 0.038228\,(\log g)^3 + 0.131524\,{\rm [Fe/H]}
\end{split}
\end{equation}
\begin{equation}
\label{eqn:TorresR}
\begin{split}
\log_{10}(R/R_{\odot}) = & ~2.793499 + 1.343062\,X + 0.723002\,X^2 \\
& -0.197217\,X^3 -0.254468\,(\log g)^2  \\
& +0.029573\,(\log g)^3 + 0.067008\,{\rm [Fe/H]}
\end{split}
\end{equation}
\begin{equation}
\label{eqn:logM}
\begin{split}
\log_{10}(M/M_{\odot}) = & ~0.764084 + 3.014102\,X + 2.272415\,X^2 \\
& -0.042698\,Y + 0.041479\,Y^2 -0.019598\,Y^3 \\
& + 0.124842\,{\rm [Fe/H]}
\end{split}
\end{equation}
\begin{equation}
\label{eqn:logR}
\begin{split}
\log_{10}(R/R_{\odot}) = & ~0.134862 + 0.391586\,X -0.370684\,Y \\
& + 0.041477\,{\rm [Fe/H]}
\end{split}
\end{equation}
These equations can be used to estimate the mass and radius of stars with T$_{\rm eff}$ and $\log g$ within the range covered by the calibration sample, as shown in Fig.~\ref{fig:Teff_logg_DEBCat}.
Most of the stars in the calibration sample have metallicities in the range $-0.4 < {\rm [Fe/H]} <+0.3$, so it is not advisable to use the relations to estimate the mass or radius of stars outside this range. 

The  root mean square (rms)  of the residuals for equations (\ref{eqn:TorresM}) and (\ref{eqn:TorresR}) are 0.021 for $\log_{10}(M/M_{\odot})$ (5.0\,per~cent) and  0.011 for $\log_{10}(R/R_{\odot})$  (2.6\,per~cent), respectively.
The rms of the residuals for equations (\ref{eqn:logM}) and (\ref{eqn:logR}) are 0.020 (4.8\,per~cent) and 0.008 (1.9\,per~cent), respectively.
For comparison, the rms of the residuals using the original coefficients from \citet{2010A&A...516A..33E} in equations (\ref{eqn:logM}) and (\ref{eqn:logR}) are 0.030 (7.2\,per~cent) and  0.014 (3.2\,per~cent), respectively.
Equation~(\ref{eqn:logM}) is the only one of these polynomial functions used in this study. 
The other polynomial functions are provided here for the convenience of other researchers interested in estimating the properties of FGK-type stars on or near the main sequence that have metallicities in the range $-0.4 < {\rm [Fe/H]} <+0.3$.

\section{Mean stellar density benchmark stars}
\label{sec:sample}

In this section, I describe the selection of benchmark stars and,
where necessary, re-analysis of the available observations to derive
accurate values of the mean stellar density, $\rho_{\star}$. 

\begin{table*}[h]
\centering
\caption{Fundamental parameters for new stellar density benchmark stars. Sources listed in the final column are as follows:
1 -- Southworth (2011); 2 -- This
work; 3 -- Agol et al. (2010); 4 -- Southworth (2010); 5 -- Southworth (2012); 6 -- Deline et al. (2025); 7 -- Bernabò et al. (2025); 8 -- Díaz et al. (2013); 9 --
Swayne et al. (2024); 10 -- Maxted et al. (2022a); 11 -- Fitzpatrick et al. (2025a); 12 -- Maxted et al. (2024); 13 -- Duck et al. (2023); 14 -- Eigmüller et al. (2018);
15 -- Rodel et al. (2024); 16 -- Triaud et al. (2013).
\label{tab:summary}
}
\begin{tabular}{lrrrrrrrr}
\hline
\hline
\noalign{\smallskip}
Star & \multicolumn{1}{l}{TIC} & \multicolumn{1}{l}{T$_{\rm eff}$ [K]} & \multicolumn{1}{l}{$M/M_{\odot}$} & \multicolumn{1}{l}{$\log g$} & \multicolumn{1}{l}{[Fe/H]} & \multicolumn{1}{l}{$\log(\rho_{\star}/\rho_{\odot})$} & 
\multicolumn{1}{l}{$\frac{d\log(\rho_{\star}/\rho_{\odot})}{dM/M_{\odot}}$} & Source \\
\hline
\noalign{\smallskip}
CoRoT$-$1$^a$   &  36352297 & $6343 \pm  74$ & $1.209 \pm 0.034$ & $ 4.348\pm 0.004$ & $ 0.04 \pm 0.03$ & $-0.177 \pm 0.002$ & -- & 1 \\
HAT$-$P$-$7     & 424865156 & $6575 \pm  34$ & $1.541 \pm 0.034$ & $ 4.028\pm 0.003$ & $ 0.28 \pm 0.02$ & $-0.709 \pm 0.001$ & -- & 2 \\
HD 189733$^a$   & 256364928 & $4875 \pm  43$ & $0.775 \pm 0.018$ & $ 4.587\pm 0.004$ & $-0.08 \pm 0.03$ & $ 0.278^{+0.005}_{-0.004}$ & -- & 3 \\
HD 209458$^a$   & 420814525 & $6126 \pm  18$ & $1.136 \pm 0.023$ & $ 4.365\pm 0.006$ & $ 0.04 \pm 0.01$ & $-0.137^{+0.007}_{-0.008}$ & -- & 4 \\
Kepler$-$1      & 399860444 & $5875 \pm  13$ & $1.035 \pm 0.020$ & $ 4.471\pm 0.004$ & $ 0.01 \pm 0.01$ & $ 0.041 \pm 0.003$ & -- & 2 \\
WASP$-$4$^a$    & 402026209 & $5496 \pm  19$ & $0.947 \pm 0.019$ & $ 4.489\pm 0.007$ & $ 0.05 \pm 0.01$ & $ 0.088 \pm 0.010$ & -- & 5 \\
WASP$-$18       & 100100827 & $6599 \pm  48$ & $1.362 \pm 0.033$ & $ 4.361\pm 0.004$ & $ 0.22 \pm 0.03$ & $-0.183 \pm 0.004$ & -- & 6 \\
WASP$-$43       &  36734222 & $4652 \pm 201$ & $0.734 \pm 0.040$ & $ 4.635\pm 0.010$ & $-0.08 \pm 0.08$ & $ 0.362^{+0.010}_{-0.008}$ & -- & 7 \\
KOI$-$205       & 271749758 & $5237 \pm  60$ & $0.905 \pm 0.038$ & $ 4.550\pm 0.015$ & $ 0.14 \pm 0.12$ & $ 0.190 \pm 0.020$ & -- & 8 \\
EBLM J0057$-$19 & 268529943 & $5580 \pm 150$ & $1.067 \pm 0.070$ & $ 4.263\pm 0.014$ & $ 0.23 \pm 0.09$ & $-0.275 \pm 0.017$ & $0.017$ & 9 \\
EBLM J0113+31 & 400048097 & $6124 \pm 50$ & $1.029 \pm 0.025$ & $4.148 \pm 0.006$ & $-0.33 \pm 0.15$ & $-0.441 \pm 0.007$ & -- & 10 \\
EBLM J0123+38   & 186514178 & $6182 \pm  91$ & $1.533 \pm 0.086$ & $ 3.932\pm 0.012$ & $ 0.45 \pm 0.07$ & $-0.852 \pm 0.016$ & $0.022$ & 9 \\
EBLM J0228+05   & 422844353 & $6912 \pm 134$ & $1.482 \pm 0.093$ & $ 4.216\pm 0.010$ & $ 0.08 \pm 0.08$ & $-0.417 \pm 0.010$ & $0.011$ & 11 \\
EBLM J0500$-$46 & 161577376 & $5788 \pm 116$ & $1.026 \pm 0.058$ & $ 4.187\pm 0.011$ & $-0.15 \pm 0.05$ & $-0.381 \pm 0.014$ & $0.022$ & 11 \\
EBLM J0526$-$34 &  24397947 & $6307 \pm 128$ & $1.332 \pm 0.080$ & $ 3.969\pm 0.009$ & $-0.05 \pm 0.07$ & $-0.765 \pm 0.008$ & $0.025$ & 11 \\
EBLM J0540$-$17 &  46627823 & $6290 \pm  77$ & $1.266 \pm 0.067$ & $ 4.072\pm 0.012$ & $-0.04 \pm 0.05$ & $-0.601 \pm 0.017$ & $0.016$ & 9 \\
EBLM J0608$-$59 & 260128333 & $6031 \pm 56$ & $1.098 \pm 0.018$ & $4.240 \pm 0.010$ & $0.01 \pm 0.05$ & $-0.322 \pm 0.016$ & -- & 12 \\
EBLM J0627$-$67 & 167201539 & $6498 \pm 117$ & $1.217 \pm 0.082$ & $ 4.255\pm 0.010$ & $-0.16 \pm 0.13$ & $-0.316 \pm 0.007$ & $0.027$ & 11 \\
EBLM J0719+25   & 458377744 & $6026 \pm  67$ & $1.142 \pm 0.059$ & $ 4.236\pm 0.011$ & $ 0.04 \pm 0.05$ & $-0.332 \pm 0.015$ & $0.017$ & 9 \\
EBLM J0941$-$31 &  25776767 & $6504 \pm 101$ & $1.412 \pm 0.081$ & $ 4.042\pm 0.009$ & $ 0.08 \pm 0.07$ & $-0.669 \pm 0.010$ & $0.017$ & 9 \\
EBLM J0955$-$39 &  45599777 & $6340 \pm  80$ & $1.097 \pm 0.061$ & $ 4.427\pm 0.009$ & $-0.24 \pm 0.08$ & $-0.037 \pm 0.010$ & $0.023$ & 9 \\
EBLM J1013+01   & 277712294 & $5200 \pm  80$ & $0.900 \pm 0.050$ & $ 4.416\pm 0.007$ & $ 0.09 \pm 0.08$ & $-0.010 \pm 0.005$ & $0.027$ & 9 \\
EBLM J1305$-$31 & 124095230 & $5913 \pm  64$ & $1.192 \pm 0.060$ & $ 4.153\pm 0.008$ & $ 0.20 \pm 0.04$ & $-0.466 \pm 0.010$ & $0.028$ & 9 \\
EBLM J1522+42   & 116380226 & $5738 \pm  64$ & $1.043 \pm 0.053$ & $ 4.175\pm 0.012$ & $-0.06 \pm 0.04$ & $-0.403 \pm 0.017$ & $0.023$ & 9 \\
EBLM J1928$-$38 & 469755925 & $5687 \pm  62$ & $1.050 \pm 0.053$ & $ 4.161\pm 0.008$ & $-0.01 \pm 0.04$ & $-0.426 \pm 0.011$ & $0.034$ & 9 \\
EBLM J1934$-$42 & 143291764 & $5648 \pm  68$ & $1.061 \pm 0.055$ & $ 4.465\pm 0.011$ & $ 0.29 \pm 0.05$ & $ 0.028 \pm 0.014$ & $0.022$ & 9 \\
EBLM J2040$-$41 & 291116020 & $5790 \pm  63$ & $1.014 \pm 0.052$ & $ 4.172\pm 0.013$ & $-0.21 \pm 0.04$ & $-0.402 \pm 0.018$ & $0.019$ & 9 \\
EBLM J2046+06   & 383303945 & $6302 \pm  70$ & $1.199 \pm 0.062$ & $ 4.287\pm 0.009$ & $ 0.00 \pm 0.05$ & $-0.266 \pm 0.011$ & $0.023$ & 9 \\
EBLM J2046$-$40 & 369456296 & $5763 \pm  75$ & $1.223 \pm 0.064$ & $ 4.086\pm 0.007$ & $ 0.34 \pm 0.05$ & $-0.572 \pm 0.007$ & $0.017$ & 9 \\
EBLM J2217$-$04 & 439837578 & $5848 \pm  85$ & $1.221 \pm 0.066$ & $ 4.096\pm 0.007$ & $ 0.27 \pm 0.06$ & $-0.556 \pm 0.007$ & $0.022$ & 13 \\
EBLM J2315+23   & 436520557 & $6027 \pm  66$ & $1.175 \pm 0.061$ & $ 4.125\pm 0.007$ & $ 0.02 \pm 0.05$ & $-0.504 \pm 0.007$ & $0.024$ & 9 \\
EBLM J2343+29   & 405532120 & $4984 \pm  87$ & $0.844 \pm 0.045$ & $ 4.544\pm 0.008$ & $ 0.11 \pm 0.05$ & $ 0.195 \pm 0.009$ & $0.020$ & 9 \\
EBLM J2359+44   & 177644756 & $6799 \pm  83$ & $1.516 \pm 0.081$ & $ 4.098\pm 0.006$ & $ 0.12 \pm 0.05$ & $-0.600 \pm 0.006$ & $0.020$ & 9 \\
EPIC 219654213  & 391137939 & $6305 \pm 110$ & $1.239 \pm 0.075$ & $ 4.109\pm 0.007$ & $-0.08 \pm 0.09$ & $-0.540 \pm 0.005$ & $0.019$ & 14 \\
NGTS$-$EB$-$7   & 238060327 & $5770 \pm 110$ & $1.177 \pm 0.076$ & $ 4.131\pm 0.015$ & $ 0.26 \pm 0.12$ & $-0.495 \pm 0.019$ & $0.011$ & 15 \\
WASP$-$30       &   9725627 & $6190 \pm  50$ & $1.201 \pm 0.065$ & $ 4.245\pm 0.019$ & $ 0.07 \pm 0.08$ & $-0.328 \pm 0.028$ & $0.005$ & 16 \\
\noalign{\smallskip}
\hline
\multicolumn{8}{@{}l}{$^a$ With correction to mean stellar density allow for eccentricity -- see section 3.1.}
\end{tabular}
\end{table*}

\subsection{Transiting exoplanets}
\label{sec:hj}

To identify transiting exoplanet systems that may be suitable as stellar density benchmark stars I used the list of systems with secondary eclipse depth measurements from \citet{2022ApJS..260....3C}.
TEPCat\footnote{\url{https://www.astro.keele.ac.uk/jkt/tepcat}} \citep{2011MNRAS.417.2166S} has also been a very useful resource to identify suitable systems for this study. 
Where possible, stellar atmospheric parameters have been taken from the SWEET-Cat catalogue of planet host star properties \citep{2021A&A...656A..53S}.\footnote{\url{https://sweetcat.iastro.pt/}}

\subsubsection{CoRoT-1\label{sec:CoRoT-1}}
Our estimate for the mean stellar density of CoRoT-1 is based on the value $\tilde{\rho} = 0.660 \pm 0.019\,\rho_{\odot}$ from \citet{2011MNRAS.417.2166S} that was computed assuming a circular orbit. 
To account for the possibility of a slightly eccentric orbit, 40 spectra of CoRoT-1 observed with the High Accuracy Radial velocity Planet Searcher (HARPS) spectrograph \citep{2000SPIE.4008..582P} were obtained from the ESO archive\footnote{\url{http://archive.eso.org}} (excluding spectra observed during the transit of CoRoT-1\,b).
These spectra were used to measure the radial velocity (RV) using cross-correlation against a numerical G2V mask with {\sc iSpec} \citep{2014A&A...569A.111B}. The posterior probability distribution (PPD) for the parameters of a Keplerian orbit fit to these RV measurements was sampled using \texttt{emcee} \citep{2013PASP..125..306F} with the time of mid-transit and orbital period fixed at the values from \citet{2022ApJS..259...62I} and a prior on $e\cos{\omega} = 0.000 \pm 0.002$ based on the phase of the secondary eclipse measured by \citet{2011ApJ...726...95D}.  
I used 1024 steps to sample the PPD after 4096 burn-in steps. 
The number of walkers is set to be 4 times the number of free parameters in the model. 
The radial velocity measurements are assumed to have a Gaussian white-noise error distribution with standard error $\sqrt{\sigma_i^2 + \sigma_{\rm jit}^2}$, where $\sigma_i$ is the precision of the radial velocity measurement obtain at time $t_i$ and $\sigma_{\rm jit}^2$ is the ``jitter'' due to stellar activity and instrumental noise. 
Convergence of the sampler was checked by visual inspection of the sample values for each parameter and each walker as a function of step number. 
The results of this analysis are given in Table~\ref{tab:CoRoT-1}.
The  density correction factor computed from this sampled PPD using equation (\ref{eqn:rhotilde}) is $\psi = 1.001 \pm 0.038$ so the mean stellar density accounting for the possibility of an eccentric orbit is $\rho_{\star} = \tilde{\rho}\times\psi = 0.663\pm0.034\rho_{\odot}$.

\subsubsection{HAT-P-7}
I have measured the mean stellar density of HAT-P-7 from the {\it Kepler} short-cadence light curve including a prior on $e\cos\omega = -0.00025 \pm 0.00078$ based on the times of secondary eclipse reported by \citet{2016ApJ...823..122W}, and a Keplerian orbit fit to a large set of radial velocity measurements.

The short-cadence pre-search data conditioning SAP fluxes provided in column PDCSAP\_FLUX of the {\it Kepler} archive files were processed quarter-by-quarter using a running-median filter to identify and remove outliers and then divided by a Gaussian process with a kernel $k(\tau) = a\,e^{-\gamma\,\tau}$ computed using {\sc celerite} to remove trends in the data due to instrumental effects and intrinsic stellar variability on time-scales $1/\gamma\sim 10$\,d with amplitude $a\sim 50$\,ppm. 

There are two companions a few arcseconds distant from HAT-P-7 detected by \citet{2017AJ....153...71F} that are more than 5 magnitudes fainter at near infrared wavelengths. 
I assume that the contribution to the optical flux from these companions is negligible so the third-light contribution was fixed at $\ell_3 = 0$ in the fit to the {\it Kepler} light curve.
A Gaussian white-noise model was used with equal standard deviations $\sigma_w$ for each observation in the {\it Kepler} light curve.
Only {\it Kepler} data observed within 160\,minutes of the predicted times of mid-transit were used for this analysis. 

I use the qpower-2 algorithm \citep{2019A+A...622A..33M} implemented in the software package {\sc pycheops} \citep{2022MNRAS.514...77M} to model the transit in the {\it Kepler} light curve assuming a power-2 limb-darkening law for the host star, i.e. the specific intensity emitted at an angle $\theta$ to the surface normal vector is 
\[I_{\lambda}(\mu) = 1-c\left(1-\mu^{\alpha}\right),\] 
where $\mu = \cos(\theta)$. The planet is assumed to be a dark sphere of radius $R_{\rm pl}$ orbiting a spherical star of radius $R_{\star}$.

Our model for the radial velocity measurements is based on a Keplerian orbit 
\[V_{K}(t) = K\left(\cos(\nu+\omega)+e\cos(\omega)\right),\] 
where $\nu$ is the true anomaly at time $t$. 
There is evidence for a third body in the HAT-P-7 system from drifts in the mean radial velocity of HAT-P-7 \citep{2009ApJ...703L..99W}.
To account for this drift, I use a model where the radial velocity measurement belonging to data set $j$ obtained at time $t_i$ is assumed to be
\[ V_r(t_i) = V_{0,j} + \dot{V}_r(t_i-T_0) + \ddot{V}_r(t-T_0)^2 +V_{K}(t_i),\]
where $T_0$ is the time of mid-transit. 
I also include a linear change in orbital period with time in the model, $\dot{P}$, to allow for the variation in the light travel time to HAT-P-7 due to its orbit around the supposed third body.
The radial velocity data sets used in the analysis are numbered as follows: 1 -- \citet[][HARPS-N]{2017A&A...602A.107B}; 2 -- \citet[][HIRES]{2008ApJ...680.1450P}; 3 -- \citet[][HIRES]{2009ApJ...703L..99W}; 4 --\citet[][SOPHIE]{2012MNRAS.422.3151H}; 5 -- \citet[][HIRES]{2014ApJ...785..126K}.
Radial velocity measurements obtained during the transit were excluded from the fit. 
The radial velocity measurements are assumed to have a Gaussian white-noise error distribution with standard error $\sqrt{\sigma_i^2 + \sigma_{\rm jit}^2}$, where $\sigma_i$ is the precision of the radial velocity measurement obtain at time $t_i$ and $\sigma_{\rm jit}^2$ is the ``jitter'' due to stellar activity and instrumental noise that is assumed to be a constant that is common to all radial velocity measurements.

The free parameters in the fit are: the time of mid-transit, $T_0$;  
the orbital period, $P$; period derivative, $\dot{P}$, the transit impact parameter, $b= a\cos(i)/R_{\star}$ where $a$ is the orbital semi-major axis and $i$ is the orbital inclination; the transit depth parameter, $D= (R_{\rm pl}/R_{\star})^2 = k^2$; the transit width parameter, $W=(R_{\star}/a)\sqrt{(1+k)^2 - b^2}/\pi$; the limb-darkening parameters $q_1$ and $q_2$ \citep{2019RNAAS...3..117S}; $f_c = \sqrt{e}\cos{\omega}$; $f_s = \sqrt{e}\sin{\omega}$; $\sigma_w$; the semi-amplitude of the spectroscopic orbit, $K$; $\dot{V}$; $\ddot{V}$; the zero points of the radial velocity data sets $V_{0,1} \dots V_{0,5}$;  the jitter on the radial velocity measurements, $\sigma_{\rm jit}$; a flux scaling factor for the {\it Kepler} light curve, $f$.  
I assume uniform priors on $\cos i$, $\log k$, $\log a/R_{\star}$.  
The limb-darkening parameters $q_1$ and $q_2$ are defined by \citet{2019RNAAS...3..117S} such that the allowed values of $ c=1-{q}_{2}\sqrt{{q}_{1}}$ and $\alpha =\log_2\left( (1-{q}_{2}\sqrt{{q}_{1}})/(1-\sqrt{{q}_{1}})\right)$ can be uniformly sampled by selecting values of $q_1$ and $q_2$ in the range 0 to 1. 
All other parameters have broad uniform priors.

\begin{figure}
    \centering
    \includegraphics[width=1\linewidth]{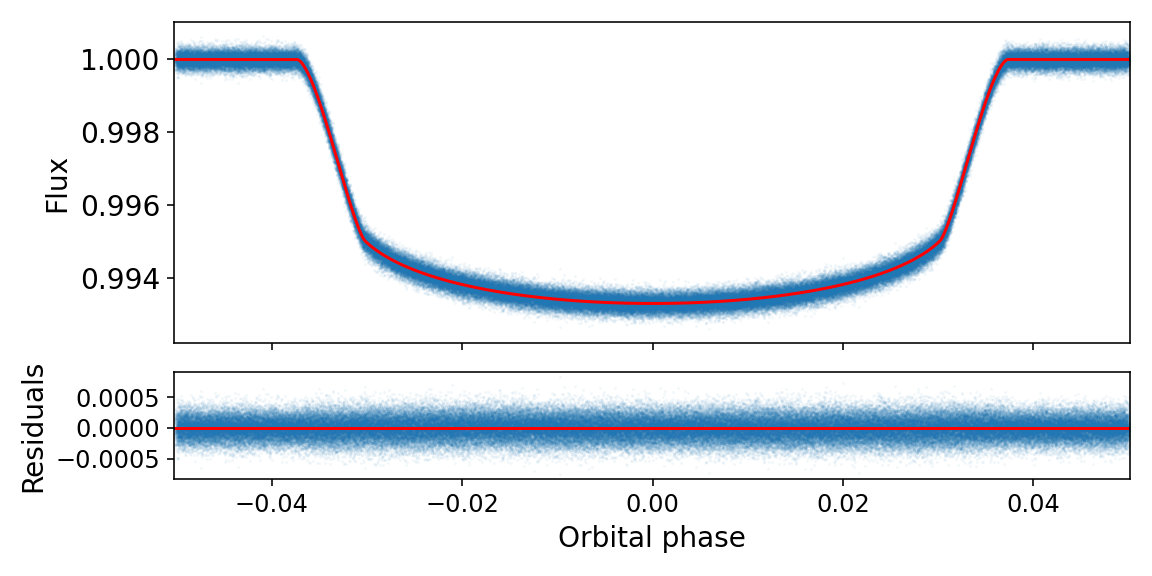}
    \includegraphics[width=1\linewidth]{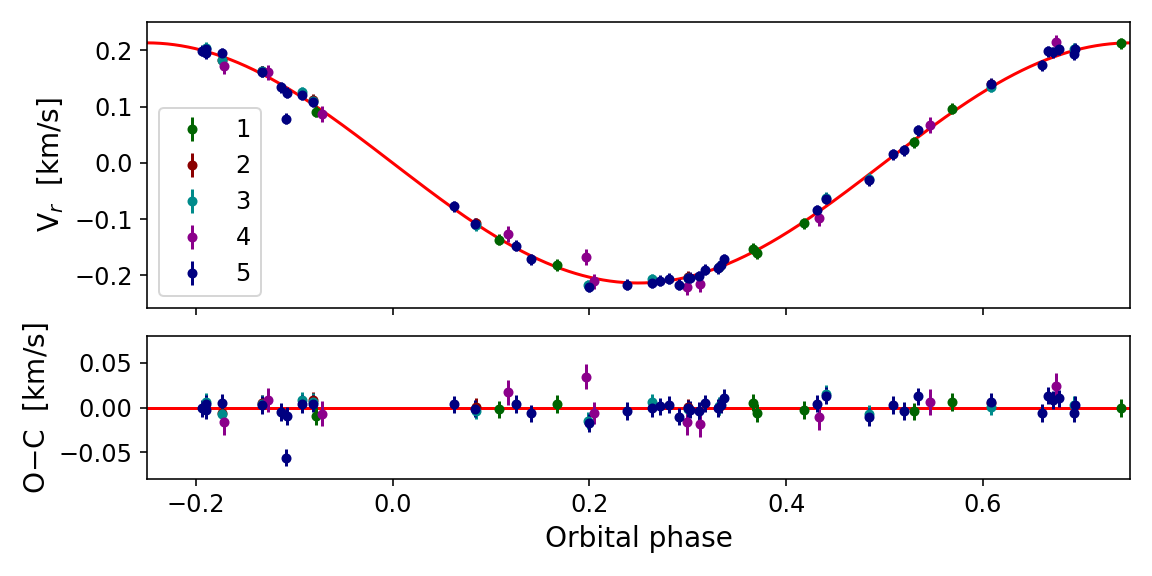}
    \caption{Upper panel: Maximum-likelihood model fit to the {\it Kepler} light curve of HAT-P-7.
    Lower panel: Maximum-likelihood model fit to radial velocity measurements. Note that systemic velocity has been subtracted from each radial velocity measurement for ease of plotting.The error bars shown here include the additional noise (``jitter'') computed as part of the analysis.}
    \label{fig:HAT-P-7_lcfit}
\end{figure}

To sample the posterior probability distribution (PPD) of the model parameters  \texttt{emcee} \citep{2013PASP..125..306F} was used with 1024 steps after 4096 burn-in steps. 
The number of walkers is set to be 4 times the number of free parameters in the model. 
Convergence of the sampler was checked by visual inspection of the sample values for each parameter and each walker as a function of step number. 

The fit to the {\it Kepler} light curve is shown in Fig.~\ref{fig:HAT-P-7_lcfit}. 
The mean and standard deviation of PPD for each of the model parameters and other details of the fit are given in Table~\ref{tab:HAT-P-7}.

\subsubsection{HD 189733}
\citet{2010ApJ...721.1861A} have derived a precise value for the mean stellar density of HD~189733 assuming a circular orbit based on their analysis of six transits of HD~189733\,b observed with Spitzer space telescope \citep{2004ApJS..154....1W} at 8 $\mu$m ($\tilde{\rho} = 1.894 \pm 0.012\,\rho_{\odot}$). 
\citeauthor{2010ApJ...721.1861A} also observed six eclipses of HD~189733\,b with the same instrument from which they derive a tight constraint on the value of $e\cos(\omega) = 0.000050 \pm 0.000094$.
To account for the possibility of a slightly eccentric orbit, the radial velocity measurements from the following sources were analysed together with this constraint on $e\cos(\omega)$: 1 -- \citet[][SOPHIE]{2009A&A...495..959B}; 2 -- \citet[][HIRES]{2006ApJ...653L..69W}; 3 -- \citet[][SPIRou]{2020A&A...642A..72M}. 
Radial velocity measurements observed during the transit were excluded from the analysis.
The values of $T_0$ and $P$ were fixed at the values measured by \citet{2023A&A...672A..24K}. 
The details of this analysis are otherwise very similar to those described for CoRoT-1 in Section~\ref{sec:CoRoT-1}.
The results are given in Table~\ref{tab:HD189733}.
The median value of $\psi$ with the 15.9\,per~cent and 84.1\,per~cent percentiles points of the PPD is given for the error estimate because these are significantly different from each other, i.e. the PPD for this parameter is significantly skewed. 
The mean stellar density with the error estimate computed in the same way is $\rho_{\star} = \tilde{\rho}\times\psi = 1.897^{+0.023}_{-0.016}\,\rho_{\odot}$. 
 
\subsubsection{HD 209458}
Our estimate for the mean stellar density of HD~209458 is based on the value $\tilde{\rho} = 0.733 \pm 0.008$ from \citet{2010MNRAS.408.1689S}  based on the analysis of several light curves obtained from space-based instrumentation described in \citet{2008MNRAS.386.1644S} and computed assuming a circular orbit. 
\citeauthor{2012ApJ...752...81C} derived a value of $e\cos(\omega) = 0.00004 \pm 0.00033$  based on their observation of the eclipse of HD~209458\,b observed at 24\,$\mu$m with Spitzer.
To account for the possibility of a slightly eccentric orbit, radial velocity measurements from the following sources have been analysed together with this constraint on $e\cos(\omega)$: 1 -- \citet[][ELODIE]{2004A&A...414..351N}; 2 -- \citet[][CORALIE]{2004A&A...414..351N};
3 -- \citet[][HIRES]{2006ApJ...646..505B}.
The values of $T_0$ and $P$ were fixed at the values measured by \citet{2014ApJ...790...53Z}. 
The details of this analysis are otherwise very similar to those described for CoRoT-1 in Section~\ref{sec:CoRoT-1}.
The results are given in Table~\ref{tab:HD209458}.
The mean stellar density allowed for the possibility of an eccentric orbit is $\rho_{\star} = 0.730^{+0.011}_{-0.014}\,\rho_{\odot}$. 

\subsubsection{Kepler-1}
Kepler-1 (TYC~3549-2811-1) is the host star to the transiting extrasolar planet TrES-2 \citep{2006ApJ...651L..61O}. 
\citet{2010ApJ...717.1084C} detected the secondary eclipse in this system using ground-based observations in the K$_{\rm s}$ band. 
The phase of the secondary eclipse implies a 3-$\sigma$ upper limits $|e\cos{\omega}| < 0.0090$. 

\begin{figure}
    \centering
    \includegraphics[width=1\linewidth]{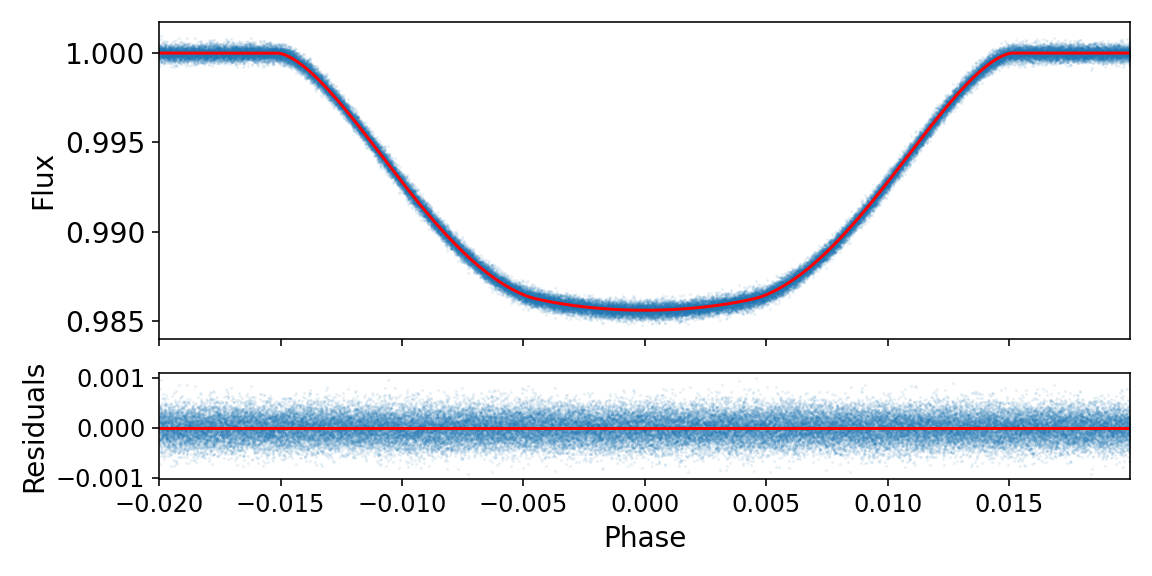}
    \includegraphics[width=1\linewidth]{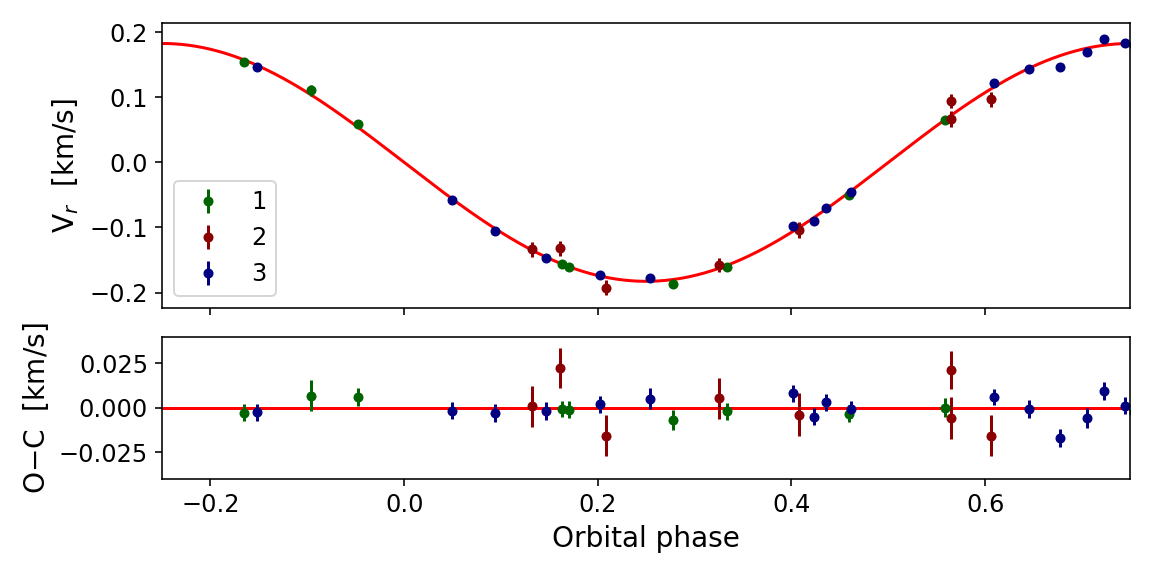}
    \caption{Upper panel: Maximum-likelihood model fit to the {\it Kepler} light curve of Kepler-1.
    Lower panel: Maximum-likelihood model fit to radial velocity measurement. Note that systemic velocity has been subtracted from each radial velocity measurement for ease of plotting. The error bars shown here include the additional noise (``jitter'') computed as part of the analysis.} 
    \label{fig:Kepler-1}
\end{figure}

To measure the mean stellar density, I analysed the {\it Kepler} light curve together with radial velocity data from the following sources: 1-- \citet[][HARPS-N]{2017A&A...602A.107B}; 2 -- \citet[][SOPHIE]{2012MNRAS.422.3151H}; 3 -- \citet[][HIRES]{2014ApJ...785..126K}.
The PDCSAP\_FLUX data for Kepler-1 were processed in the same way as described above for HAT-P-7 prior to analysis.
The joint fit to the {\it Kepler} light curve and the radial velocity data were also similar to that described above for HAT-P-7 except that third light is included as a free parameter in the fit with a Gaussian prior $\ell_3 = 0.020\pm0.005$.  
This prior on $\ell_3$ is based on the detection of a companion star $1.1$\,arcsec from Kepler-1 that is approximately 4.3 magnitudes fainter in the {\it Kepler} band pass \citep{2017AJ....153...71F}. 
A Gaussian prior on  $e\cos{\omega} =  0.000 \pm 0.003$ was used based on the 3-$\sigma$ upper limit from \citeauthor{2010ApJ...717.1084C}. 
No linear drift in the radial velocity data was included in the fit and that the orbital period is assumed to be constant.
Only {\it Kepler} data observed within 71\,minutes (transit width / 1.5) of the predicted times of mid-transit has been used for the analysis. 
The fit to the {\it Kepler} light curve is shown in Fig.~\ref{fig:Kepler-1}. 
The results of the fit are given in Table~\ref{tab:Kepler-1}. 

\subsubsection{WASP-4}
The estimate for the mean stellar density of WASP-4 is based on the value $\tilde{\rho} = 1.230 \pm 0.022$ from \citet{2012MNRAS.426.1291S} which is taken from the analysis of high-quality ground-based light curves analysed by \citet{2009MNRAS.399..287S} assuming a circular orbit. 
\citeauthor{2011ApJ...727...23B} derived a value of $e\cos(\omega) = 0.00030 \pm 0.00086$  based on their observation of the eclipse of WASP~4\,b observed at 3.6 and 4.5\,$\mu$m with Spitzer.
Radial velocity measurements from the following sources have been analysed together with this constraint on $e\cos(\omega)$: 1 -- \citet[][HIRES]{2014ApJ...785..126K}; 2 -- \citet[][CORALIE]{2010A&A...524A..25T};
3 -- \citet[][HARPS]{2010A&A...524A..25T}.
The values of $T_0$ and $P$ were fixed at the values measured by \citet{2022ApJS..259...62I}. 
The deviation in the time of mid-transit from this linear ephemeris due to the slow drift in orbital measured by \citeauthor {2022ApJS..259...62I} ($\dot{P} = -5.8 ± 1.6$\,ms\,yr$^{-1}$) is less than 1\, minute over 10 years and so was ignored for this analysis. 
The details of this analysis are otherwise very similar to those described for CoRoT-1 in Section~\ref{sec:CoRoT-1}.
The results are given in Table~\ref{tab:WASP-4}.
The correction for the eccentricity of the orbit is $\psi = 0.999 \pm 0.012$, so the mean stellar density is found to be $\rho_{\star} = 1.225\pm0.029\,\rho_{\odot}$.

\subsubsection{WASP-18}
WASP-18 is a short-period system (P=0.9415\,d) in which a star with a mass $\approx 1.3\,M_{\odot}$ is transited by an ultra-hot Jupiter companion with a mass $\approx 10.5\,M_{\rm Jup}$.
\citet{deline2025} have measured the value $a/R_{\star} = 3.493 \pm 0.011$ from a simultaneous fit to light curves observed with CHEOPS \citep{2021ExA....51..109B}, Spitzer and TESS \citep{Ricker14}. 
The simultaneous fit of these multiple data sets, each with its own sources of systematic noise, requires a complex model with over 100 parameters, so Deline et al. assumed a circular orbit for this analysis. 
Nevertheless, they find that the orbit of WASP-18\,b is slightly eccentric, with $e = 0.00852 \pm 0.00091$ and $\omega = 261.9\degr \pm 1.4\degr$.  
This is based mostly on the phase of the secondary eclipse observed with the NIRISS instrument on the James Webb space telescope (JWST) by \citet{2023Natur.620..292C}. 
The correction for the eccentricity of the orbit is $\psi = 1.0256 \pm 0.0028$, so the mean stellar density is $\rho_{\star} = 0.6566 \pm 0.0065 \rho_{\odot}$.

\subsubsection{WASP-43}
WASP-43 is a K7\,V star that has a hot-Jupiter companion with a very short orbital period (P=0.81347\,d). 
\citet{2025A&A...694A.233B} have published a detailed study of the tidal decay and apsidal motion in this system based on the full-phase light curve acquired by JWST with the MIRI instrument between 6.5 and 7.0\,$\mu$m, two sectors of TESS data, new and archival radial velocity measurements, and published times of mid-transit and mid-eclipse. 
This analysis reveals a small but significant eccentricity in the orbit of WASP-43\,b ($e=0.00188\pm0.00035$). 
There are three values of $a/R_{\star}$ quoted in this study, one from the analysis of the JWST lightcurves and one from each of the two sectors of TESS data. 
All these values are based on an eccentric orbit fit (L.~Bernab{\`o}, priv. comm.) and agree well with each other. 
Using the weighted mean value $a/R_{\star} = 4.845 \pm 0.028$ the mean stellar density of WASP-43 is found to be $\rho_{\star} = 2.300\pm 0.040\,\rho_{\odot}$.

\subsection{Eclipsing binaries with directly measured mass ratios}
There are currently only two eclipsing binaries (EBs) with flux ratios $\lesssim$\,0.2\,\% at optical wavelengths that have published mass ratio values based on direct measurements using near-infrared spectroscopy -- EBLM~J0113+31 \citep{2022MNRAS.513.6042M} and EBLM~J0608$-$59 \citep{2024MNRAS.531.4577M}. 
The primary stars of these (EBs) are both useful benchmark stars for stellar density, and also for stellar mass, radius, luminosity, etc. 
The apparent magnitudes and flux ratio at several optical and near-infrared wavelengths for these EBs are provided in the \citet{2022MNRAS.513.6042M} and \citet{2024MNRAS.531.4577M}.

\subsection{EBLM and transiting brown dwarf systems}
\label{sec:eblm}
Eclipsing binaries with very low mass companions (EBLM systems)  can also be used as stellar density benchmark stars. 
In this section, I describe the selection of suitable EBLM systems and how I have used the empirical $M_{\star}(\rho_{\star}, {\rm T}_{\rm eff}, {\rm [Fe/H]})$  relation derived in Section~\ref{sec:recalib} to infer the value of the mean stellar density, $\rho_{\star}$, for the primary stars in these binary systems.

The majority of the measurements of $R_{\star}/a$ for EBLM systems in the sample are taken from \citet{2024MNRAS.528.5703S} and \citet{2025MNRAS.tmp..257F}. 
All systems where the estimated mass ratio is $q>0.3$ have been excluded from the sample.
Useful measurements of $R_{\star}/a$ for individual EBLM systems have been taken from \citet{2018MNRAS.480.3864E} (EPIC~219654213) and \citet{2023MNRAS.521.6305D} (EBLM~J2217-04). 
The transiting brown dwarf systems NGTS-EB-7 \citep{2024MNRAS.tmp.2675R} and WASP-30 \citep{2013A&A...549A..18T} have also been included in the sample. 
Measurements of the orbital eccentricity, $e$, orbital inclination, $i$, orbital period, $P$, the semi-amplitude of the primary star's spectroscopic orbit, $K_1$, and estimates of T$_{\rm eff}$ and [Fe/H] for the primary star have been taken from the same sources.

I have used the following Monte Carlo calculation with a sample size $N=100\,000$  to find an estimate of $M_{\star}$ that is consistent with the mass estimate calculated using equation~(\ref{eqn:logM}) and the standard error on the parameters of interest. 
An iterative approach is used here because the mean stellar density that is used to estimate  $M_{\star}$  has a weak dependence on the assumed mass ratio, $q$, which is itself computed from the mass function\footnote{$f(m) = (M_2 \sin i)^3/(M_{\star}+M_2)^2 = 1/(2\pi G)K_1^3 P$} using an estimate of  $M_{\star}$. 

\begin{enumerate}
\item Generate $N$ sets of random values for $M_{\star}$, $R_{\star}/a$, $e$, $i$, $K_1$, T$_{\rm eff}$ and [Fe/H] from normal distributions, each with the mean and standard deviation equal to the published values and standard errors on these parameters.
\item Calculate the mass ratio, $q$, for all $N$ sets of $P$, $e$, $i$, $K_1$ and $M_{\star}$ values in the random sample. 
\item Calculate $N$ values of $\log(\rho_{\star}/\rho_{\odot})$ from the random samples of $R_{\star}/a$ and $q$ values using equation (\ref{eqn:rhostar}).
\item Use the empirical polynomial function in equation~(\ref{eqn:logM}) to calculate $N$ values of $\log(M_{\star}/M_{\odot})$.
\item Calculate $N$ random values with a normal distribution with a mean of 0 and a standard deviation of 0.020. Add these to the random sample of $\log(M_{\star}/M_{\odot})$ values from the previous step to account for the observed scatter around this relation.
\item Compute a random sample of $M_{\star}$ values from 
the random sample of $\log(M_{\star}/M_{\odot})$ values.
\item Repeat from step (ii) two more times.
\end{enumerate}
The convergence of this algorithm was checked by calculating the change in $\rho_{\star}$ between the last two iterations of the loop. 
This change is found to be less than 0.05\,per~cent for all stars.

The 100\,000 random values from the final iteration of this algorithm are then used to calculate the values of $\log(\rho_{\star}/\rho_{\odot})$,  $M_{\star}/M_{\odot}$ and $\log g$ and their standard errors from the mean and standard deviation of the Monte Carlo samples. 
The same random values are also used to calculate $d\log(\rho_{\star}/\rho_{\odot})/(dM/M_{\odot})$, the correlation coefficient between $\log(\rho_{\star}/\rho_{\odot})$ and $dM/M_{\odot}$. 

\begin{figure}
    \centering
    \includegraphics[width=1\linewidth]{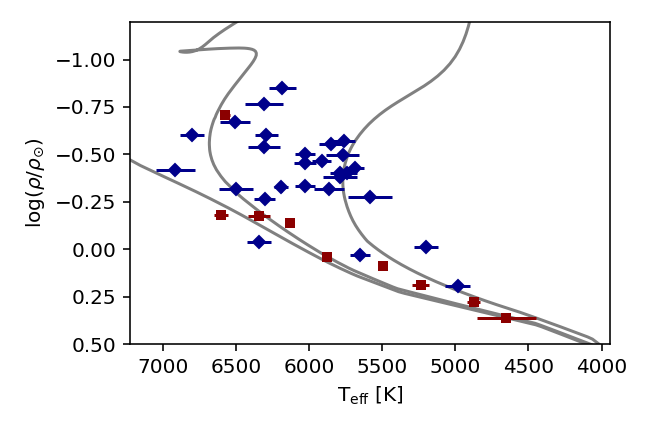}
    \caption{Distribution of planet host stars (squares) and stars in EBLM systems (diamonds) from Table~\ref{tab:summary} in the T$_{\rm eff}$\,--\,$\log (\rho/\rho_{\odot})$  plane. Lines are isochrones for solar composition at ages of 1, 5 and 10 Gyr in that order going from bottom to top \citep{2016ApJS..222....8D, 2016ApJ...823..102C}}
    \label{fig:teff_logrho_benchmarks}
\end{figure}

\subsection{Summary of results}
The values of the mass, surface gravity and mean stellar density for these benchmark stars are summarised in Table~\ref{tab:summary}.
The distribution of these benchmark stars in the T$_{\rm eff}$\,--\,$\log (\rho/\rho_{\odot})$  plane is compared to the distributions of planet hosts from SWEET-Cat and the calibration sample from DEBCat in Fig.~\ref{fig:teff_logrho_benchmarks}. It can be seen that the calibration sample covers the parameter space of the benchmark sample effectively, and that the benchmark sample is well matched to the properties of planet host stars. 

\section{Discussion}
\subsection{Distribution of benchmark star properties}
It is noticeable that the EBLM systems in Fig.~\ref{fig:teff_logrho_benchmarks} tend to be found near the main-sequence turn off, whereas all but one of the hot-Jupiter systems are close to the main sequence. 
There are obviously several selection effects at play here given that this is a sample of stars that have been selected based on the availability of high quality data and the possibility to measure an accurate value of the mean stellar density. 
There may also be some impact here from the finite lifetime of hot-Jupiter exoplanets due to tidal decay of the orbit leading to the engulfment of the planet by its host star \citep{2009ApJ...692L...9L,2019AJ....158..190H}.
Between the two samples, there is good coverage of the stellar properties of F- and G-type planet host stars in the Kiel diagram (Fig.~\ref{fig:Teff_logg_DEBCat}) but there are only two K-type stars in the sample. 
There is also good coverage of the range of [Fe/H] values down to ${\rm [Fe/H]}\approx -0.3$. 

\label{sec:discuss}
\subsection{Test of method for EBLM systems}
EBLM~J0113+31 and EBLM~J0608$-$59 are useful test cases for the method to derive the mass and mean stellar density of solar-type stars in EBLM systems that is described in Section~\ref{sec:eblm}. 
Despite the extreme flux ratio for these EBLM systems ($\approx 0.1$\,per~~cent at optical wavelengths), extensive high-quality spectroscopy at near-infrared wavelengths has made it possible to directly measure the mass ratio for these binary systems.
This makes it possible to measure the mass, radius, and mean stellar density for both stars in these systems without using stellar models or making assumptions about the properties of the primary star, i.e. without using empirical relations such as Equation~(\ref{eqn:logM}).
These model-independent values of the mass and $\log (\rho/\rho_{\odot})$ are given in Table~\ref{tab:summary} for EBLM~J0113+31 and EBLM~J0608$-$59, and can be compared to the values derived using the method described in Section~\ref{sec:eblm}.

For EBLM~J0113+31, the input values of 
T$_{\rm eff} = 6025 \pm  76$\,K, 
$\rm {[Fe/H]} =-0.31 \pm 0.05$ and 
$R_{\star}/a = 0.0539 \pm 0.0003$ 
taken from \citet{2024MNRAS.528.5703S} result in a value of 
$\log(\rho_{\star}/\rho_{\odot}) = -0.454 \pm 0.007$ and 
$M_{\star} = 1.06 \pm 0.06\,M_{\odot}$ using the method described in Section~\ref{sec:eblm}.
Similarly for EBLM~J0608$-$59, the input values of  
T$_{\rm eff} = 5865 \pm 103$\,K, 
$\rm {[Fe/H]} =+0.01 \pm 0.05$ and 
$R_{\star}/a = 0.0467 \pm 0.0002$ taken from \citet{Fitzpatrick} result in a values of 
$\log(\rho_{\star}/\rho_{\odot}) = -0.317 \pm 0.006$ and 
$M_{\star} = 1.08 \pm 0.06\,M_{\odot}$.
The agreement between these mass estimates and the values in Table~\ref{tab:summary} is excellent.
The chi-squared statistic for the difference between these two values of  $\log(\rho_{\star}/\rho_{\odot})$ and the values of $\log(\rho_{\star}/\rho_{\odot})$ in Table~\ref{tab:summary} is $\chi^2 = 2.015$ with 2 degrees of freedom.
This demonstrates that the values of the stellar mass and $\log(\rho_{\star}/\rho_{\odot})$ and their standard error estimates derived using the method described in Section~\ref{sec:eblm} are reliable. 

\subsection{Impact of flux from companion in EBLM systems}
\label{sec:lrat}
 In this section, I demonstrate that the flux from the low-mass companion in the EBLM systems is not a significant source of systematic error in the stellar properties derived from the analysis of the stellar spectrum or from the system luminosity.

Table~\ref{tab:lrat} lists the flux ratio measured from the depth of the secondary eclipse for each of the EBLM systems in the sample. 
The CHEOPS transmission function is similar to the Gaia G band \citep{2024A&A...687A.302F,2016A&A...595A...1G} so the flux ratio measured from CHEOPS light curves is used directly to estimate the flux ratio in the G band, $\ell_G$. 
To estimate  the flux ratio in the V band, $\ell_{\rm V}$, from the flux ratio measured from {\it Kepler} light curves, $\ell_{\rm Kp}$, I have used version 2019.3.22 of the colour -- effective temperature sequence for dwarf stars compiled from \citet{2013ApJS..208....9P} downloaded from Eric Mamajek's web site.\footnote{\url{http://www.pas.rochester.edu/~emamajek/EEM_dwarf_UBVIJHK_colors_Teff.txt}} 
Linear interpolation in this sequence is used to estimate (B$-$V) from T$_{\rm eff}$ for the primary star and from the mass for secondary star.
Equations (2a), (6a) and (6b) from \citet{2011AJ....142..112B} are then used to calculate $({\rm V}-{\rm Kp}) \approx 0.346 ({\rm B}-{\rm V}) -0.115$ for each star, ignoring that the small difference between the definitions of the Johnson and Tycho B and V bands.
Equation (1) from \citet{2019AJ....158..138S} is used to calculate the (G$-$T) colour from the interpolated values of (BP$-$RP) and then use the interpolated value of to estimate (G$-$V) to estimate $\ell_{\rm V}$ from the flux ratio measured from TESS light curves, $\ell_{\rm T}$.
The luminosity of the primary star, $L_1$, is calculated from $T_{\rm eff}$ and the value of R$_{\star}$ computed from the values of $M_{\star}$ and $\log(\rho_{\star}/\rho_{\odot})$.
To estimate the luminosity ratio, $L_2/L_1$, I use the mass ratio, $M_2/M_1$, derived as part of the analysis described in Section~\ref{sec:eblm} and the value of $M_{\star}$ given in Table~\ref{tab:summary} to estimate the mass of the companion star, $M_2$. Interpolation in the sequence from \citeauthor{2013ApJS..208....9P} is then used to find the luminosity of the companion, $L_2$, based on its mass, except for EBLM~J0113+31 and EBLM~J0608$-$59, where the more accurate values of $L_1$ and $L_2$ from  \citet{2022MNRAS.513.6042M} and \citet{2024MNRAS.531.4577M} are used to compute the values of $L_2/L_1$ given in Table~\ref{tab:lrat}. 

The median value of the V-band flux ratios in Table~\ref{tab:lrat}  is $\ell_{\rm V} = 0.025$\,per~cent and $\ell_{\rm V} < 0.1$\,per~cent in all cases, so the companion will have a negligible impact on estimates of the primary star's properties based on data obtain at optical wavelengths, e.g. Gaia data. 
\citet{2018MNRAS.473.5043E} have used simulated data to assess the impact of unresolved binary stars on stellar parameters from the APOGEE survey \citep{2015AJ....150..148H} based on spectra observed in the H band.
The median value of the flux ratio in the H band for the EBLM systems listed in Table~\ref{tab:lrat} is $\ell_{\rm H} = 0.8$\,per~cent and $\ell_{\rm H} < 2$\,per~cent for all these EBLM systems. Based on the inset to Fig.~1 of \citeauthor{2018MNRAS.473.5043E}, this level of flux contamination also has a negligible impact on the values of  measured from APOGEE spectra.

I used {\sc bagemass} \citep{2015A&A...575A..36M} to estimate $\log(\rho_{\star}/\rho_{\odot})$ for EBLM~J0113+31 and EBLM~J0608$-$59 from stellar isochrones computed with the {\sc garstec} stellar evolution code \citep{2008Ap&SS.316...99W} based on the observed values of $\log L/L_{\odot}$, T$_{\rm eff}$ and [Fe/H] taken from 
\citet{2022MNRAS.513.6042M} and \citet{2024MNRAS.531.4577M}. 
This was done for two cases for each star, one using the actual value of $\log L/L_{\odot}$ for the primary star and again with the total luminosity of the binary system used instead of $\log L/L_{\odot}$. 
For EBLM~J0113+31, adding the flux of the companion changes the stellar density estimate from $\log(\rho_{\star}/\rho_{\odot})=-0.439$ to $\log(\rho_{\star}/\rho_{\odot})=-0.441$, a difference of $-0.4$\,per~cent. 
For EBLM~J0608$-$59, the stellar density estimate changes from $\log(\rho_{\star}/\rho_{\odot})=-0.314$ to $\log(\rho_{\star}/\rho_{\odot})=-0.311$, a difference of $+0.6$\,per~cent.
EBLM~J0608$-$59 has one of the largest values of $L_2/L_1$ for the selected EBLM systems and the value of $L_2/L_1$ for EBLM~J0113+31 is close to the median value for this sample. 
The bias in $\log(\rho_{\star}/\rho_{\odot}) \approx \pm 0.5$\,per~cent estimated using {\sc bagemass} is small compared to the typical error on these values (2.3\,per~cent). 

\begin{table}
\centering
\caption{Flux and luminosity ratios for EBLM systems. The flux ratios measrured in the CHEOPS, {\it Kepler} and TESS bands are denoted $\ell_{\rm G}$, $\ell_{\it Kepler}$ and $\ell_{\rm T}$, respectively. The flux ratio in the V band $\ell_{\rm V}$ and the luminisity ratio $\frac{L_2}{L_1}$ have been estimated using the method described in Section~\ref{sec:lrat}.  }
\label{tab:lrat}
\begin{tabular}{lrrrrrr}
\hline\hline
 & & \multicolumn{3}{c}{Measured [\%]}  & \multicolumn{2}{c}{Estimated [\%]} \\
Star & $\frac{M_2}{M_{\star}}$ & 
\multicolumn{1}{c}{$\ell_{\rm G}$ } & 
\multicolumn{1}{c}{$\ell_{\rm Kp}$} & 
\multicolumn{1}{c}{$\ell_{\rm T}$}
& \multicolumn{1}{c}{$\ell_{\rm V}$ } & 
\multicolumn{1}{c}{$\frac{L_2}{L_1}$} \\
\hline
EBLM J0057$-$19  & 0.13 &     0.04 &          &     0.05 &     0.01 &     0.15 \\
EBLM J0113+31    & 0.19 &     0.08 &          &     0.16 &     0.02 &     0.21 \\
EBLM J0123+38    & 0.26 &     0.10 &          &     0.26 &     0.05 &     0.33 \\
EBLM J0228+05    & 0.12 &          &          &     0.06 &     0.01 &     0.09 \\
EBLM J0500$-$46  & 0.16 &          &          &     0.12 &     0.01 &     0.19 \\
EBLM J0526$-$34  & 0.25 &          &          &     0.21 &     0.04 &     0.25 \\
EBLM J0540$-$17  & 0.14 &     0.04 &          &     0.07 &     0.01 &     0.11 \\
EBLM J0608$-$59  & 0.28 &          &          &     0.41 &     0.08 &     0.44 \\
EBLM J0627$-$67  & 0.25 &          &          &     0.15 &     0.03 &     0.37 \\
EBLM J0719+25    & 0.14 &     0.06 &          &     0.09 &     0.02 &     0.16 \\
EBLM J0941$-$31  & 0.17 &     0.06 &          &     0.12 &     0.02 &     0.13 \\
EBLM J0955$-$39  & 0.19 &     0.15 &          &     0.28 &     0.05 &     0.36 \\
EBLM J1013+01    & 0.18 &     0.16 &          &     0.29 &     0.04 &     0.50 \\
EBLM J1305$-$31  & 0.25 &     0.26 &          &     0.10 &     0.10 &     0.45 \\
EBLM J1522+42    & 0.16 &     0.05 &          &     0.11 &     0.01 &     0.22 \\
EBLM J1928$-$38  & 0.27 &     0.13 &          &          &     0.05 &     0.52 \\
EBLM J1934$-$42  & 0.18 &     0.13 &          &     0.21 &     0.04 &     0.57 \\
EBLM J2040$-$41  & 0.15 &     0.03 &          &     0.09 &     0.01 &     0.15 \\
EBLM J2046+06    & 0.20 &     0.03 &          &     0.07 &     0.01 &     0.30 \\
EBLM J2046$-$40  & 0.15 &     0.09 &          &     0.19 &     0.02 &     0.19 \\
EBLM J2217$-$04  & 0.20 &          &     0.07 &          &     0.05 &     0.27 \\
EBLM J2315+23    & 0.21 &     0.10 &          &     0.17 &     0.03 &     0.27 \\
EBLM J2343+29    & 0.11 &     0.03 &          &          &     0.00 &     0.22 \\
EBLM J2359+44    & 0.22 &     0.09 &          &     0.20 &     0.03 &     0.21 \\
EPIC 219654213   & 0.17 &          &     0.05 &          &     0.03 &     0.16 \\
NGTS$-$EB$-$7    & 0.09 &          &          & $<0.037$ &     0.00 &     0.04 \\
WASP$-$30        & 0.05 &          &          &     0.01 &     0.00 &     0.01 \\
\hline
\end{tabular}
\end{table}

\subsection{Example applications}
 
The values of $\log(\rho_{\star}/\rho_{\odot})$ and  $\log g$  from Table~\ref{tab:summary} can be used directly to test the accuracy and precision of estimates for these quantities if the mass estimates from Table~\ref{tab:summary} are thought to be the best available.
If more accurate or more precise mass estimates are available then a small correction should be applied to the values of  $\log g$ and  $\log(\rho_{\star}/\rho_{\odot})$ to account for the weak covariance between the values of  $(\rho_{\star}/\rho_{\odot})$ and the mass estimate.

The value of $\log(\rho_{\star}/\rho_{\odot})$ that should be compared to the value of $\log(\rho_{\star}/\rho_{\odot})_{\rm C}$ given or derived from stellar parameter estimates in a catalogue that is consistent with the mass estimate $(M/M_{\odot})_{\rm C}$ from the same catalogue is
\begin{equation}
\label{eqn:dlogrhostar}
\log(\rho_{\star}/\rho_{\odot})_{M} =   \log(\rho_{\star}/\rho_{\odot})_{1} 
+ \frac{d\log(\rho_{\star}/\rho_{\odot})}{dM/M_{\odot}}\times \Delta{M},
\end{equation}
where $(M/M_{\odot})_{1}$, $\log(\rho_{\star}/\rho_{\odot})_{1}$, ${d\log(\rho_{\star}/\rho_{\odot})}/(dM/M_{\odot})$  are the values taken from Table~\ref{tab:summary} and  $\Delta{M} =(M/M_{\odot})_{\rm C} - (M/M_{\odot})_{1}$.
Similarly, the corrected value of the surface gravity that is consistent with the value of $(M/M_{\odot})_{\rm C}$ can be computed using the value of $(\log g)_1$ from Table~\ref{tab:summary} and the following equation:
\begin{equation}
\label{eqn:dlogg}
(\log g)_{\rm M} =  (\log g)_1 + \left(\frac{1}{3(M/M_{\odot})_{1}} +  \frac{2}{3} \frac{d\log(\rho_{\star}/\rho_{\odot})} {dM/M_{\odot}}\right) \times \Delta\,M.
\end{equation}

A comparison between the mass, mean stellar density and surface gravity values from Table~\ref{tab:summary} to the estimates of these quantities obtained from various catalogues is summarised in Table~\ref{tab:examples}.  
For each quantity I use the {\tt combine} function described in the appendix to \citet{2022MNRAS.514...77M} to compute a weighted mean offset allowing for additional scatter $\sigma_x$ in the measurements that is added in quadrature to the quoted standard error. 
The corrections given by equations (\ref{eqn:dlogrhostar}) and (\ref{eqn:dlogg}) has been applied so that all the comparisons are done against values of $\log(\rho_{\star}/\rho_{\odot})$ and $\log g$ that are consistent with the mass estimates given in each catalogue. 
The standard errors on $\log(\rho_{\star}/\rho_{\odot})$ and  $\log g$ from  Table~\ref{tab:summary} are assumed to be unaffected by this small correction.
No value of $d\log(\rho_{\star})/dM_{\star}$  is given for EBLM~J0113+31  or EBLM~J0608$-$59 because the mass ratio has been measured directly for these two binary systems, so they are not subject to this small additional source of uncertainty. 

\begin{figure*}
    \centering
    \includegraphics[width=0.45\linewidth]{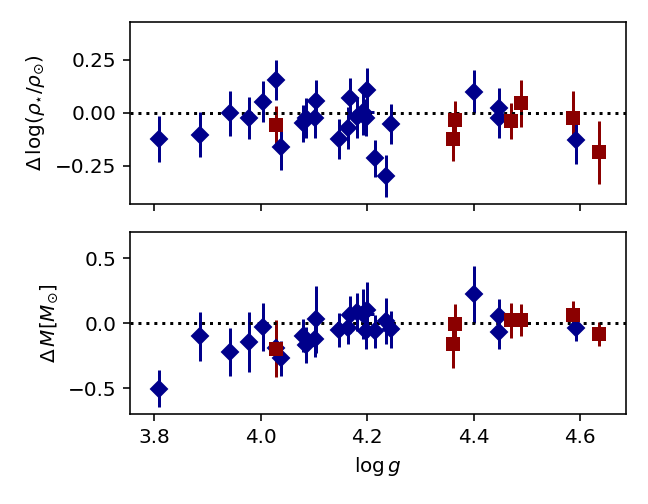}
    \includegraphics[width=0.45\linewidth]{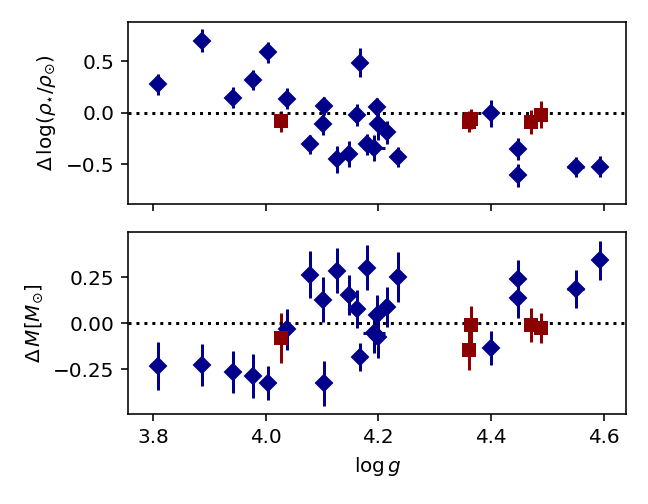}
    \caption{Left panel: Offsets relative to values from Table~\ref{tab:summary} for  mass and mean stellar density estimates from TIC~v8.2 as a function of $\log g$ values from Table~\ref{tab:summary}. Symbols as in Fig.~\ref{fig:teff_logrho_benchmarks}.
    Right panel: Same as the left panel but for mass and mean stellar density values from asPIC1.1. }
    \label{fig:delta_rho}
\end{figure*}

Version 8.2 of the TESS input catalogue \citep[TIC v8.2, ][]{2019AJ....158..138S} provides estimates for the mass, $\log g$ and $\log(\rho_{\star})$ with standard error estimates for 32 of the stars in Table~\ref{tab:summary}. 
\citeauthor{2019AJ....158..138S} state that they can estimate masses for stars
``{\it on the main sequence or not too far evolved from it. }'' but note that this includes ``{\it stars that are subgiants; these should be regarded with caution.}''. All the stars in Table~\ref{tab:summary} are well below the upper limit in radius that \citeauthor{2019AJ....158..138S} use to separate main-sequence stars and subgiants from red giants. Nevertheless, there is a small ($\approx 10$\,per~cent) but statistically significant offset in mass compared to the benchmark values in  Table~\ref{tab:summary}. It is tempting to blame this on stars that have evolved away from the main sequence as there appears to be a weak trend with surface gravity for stars with $\log g \lesssim 4.0$ visible in Fig.~\ref{fig:delta_rho}. However, there is no such trend in $\log(\rho_{\star})$ apparent in the same figure so there may be a confounding variable at play here that is responsible for these offsets. In any case, the standard error estimates for both mass and  $\log(\rho_{\star})$ in TIC v8.2 appear to be a reliable estimate of the precision of these estimates.

\begin{figure*}
    \centering
    \includegraphics[width=0.45\linewidth]{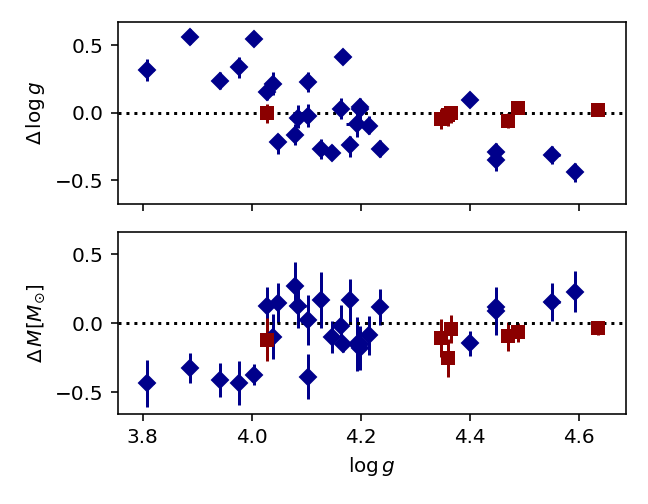}
    \includegraphics[width=0.45\linewidth]{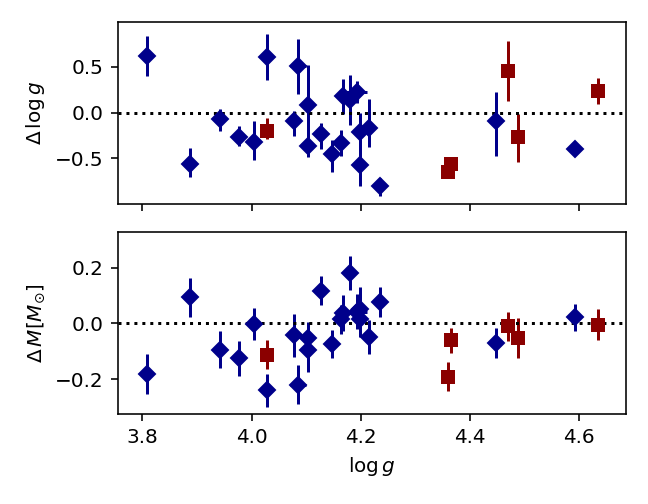}
    \caption{Left panel: Offsets relative to values from Table~\ref{tab:summary} for mass and $\log g$ estimates from the StarHorse~2 catalogue as a function of $\log g$ values from Table~\ref{tab:summary}. Symbols as in Fig.~\ref{fig:teff_logrho_benchmarks}.
    Right panel: Same as the left panel but for mass and $\log g$ estimates ({\tt logg\_gspspec\_ann}) from Gaia DR3. }
    \label{fig:delta_logg}
\end{figure*}

The first public release of the all-sky PLATO input 
catalogue (asPIC1.1) is described in \citet{2021A&A...653A..98M}. 
The StarHorse~2 catalogue is described in \citet{2019A&A...628A..94A}.
Both these catalogue are based on the Gaia DR3 catalogues and other data sources but each use their own algorithms to estimate the mass and radius of the stars.
There is no significant offset between the mass or mean stellar density estimates from asPIC1.1 and the benchmark values but there is appreciable additional scatter in both of these quantities, i.e. the standard error estimates in asPIC1.1 appear to be underestimated. It can be seen in Fig.~\ref{fig:delta_rho} that this additional scatter may be explained by a positive trend in the estimated mass as a function of $\log g $. A very similar effect is seen in Fig.~\ref{fig:delta_logg} for mass and surface gravity estimate taken from the StarHorse~2 catalogue. 

For the Gaia~DR3 catalogue \citep{2023A&A...674A...1G} I have compared  $\log(\rho_{\star}/\rho_{\odot})$ to the value computed from the mass and radius calculated with the Final Luminosity Age Mass Estimator \citep[FLAME, ][]{2023A&A...674A..26C}.
There is good agreement between these mass and surface gravity estimates and the benchmark values from Table~\ref{tab:summary} but the standard error estimates on the radius from FLAME result appear to be underestimated by about 15\,per~cent resulting in additional scatter of about 45\,per~cent in values of $\log(\rho_{\star}/\rho_{\odot})$.  
The value of $\log g$ obtained from an analysis of spectra obtained with the 
Radial Velocity Spectrometer (RVS) performed by the General Stellar Parametriser-spectroscopy \citep[GSP-Spec][]{2023A&A...674A..29R} ({\tt logg\_gspspec\_ann}) shows both a significant offset and additional scatter compared to the values in Table~\ref{tab:summary}.
This additional scatter can be seen in Fig.~\ref{fig:delta_logg}.

\begin{table*}
\centering
\caption{Summary of comparisons between mass, surface gravity and mean stellar density estimates from various catalogues to the values in Table~\ref{tab:summary}. The number of stars from each catalogue found in Table~\ref{tab:summary}, $n$, is given in the first column. 
Figures in parentheses give the offset in $(\rho_{\star}/\rho_{\odot})$ as a percentage.
The statistical significance of the offset for a two-tailed test is given in the column headed $p$. 
The quantities used to calculate $\log(\rho_{\star}/\rho_{\odot})$ and $M_{\star}$  are noted in the final column using the column names as specified by the catalogue named in the first column.}
\label{tab:examples}
\begin{tabular}{llr@{~~}rrrrl}
\hline\hline
Catalogue & Parameter & \multicolumn{2}{l}{Offset} &  \multicolumn{1}{l}{$p$}& \multicolumn{2}{l}{$\sigma_x$} & Notes\\
\hline
\noalign{\smallskip}
TIC v8.2     & $\log(\rho_{\star}/\rho_{\odot})$ & $-0.040 \pm 0.018 $ & $(-9\%)$ &  0.01 & $0.004 \pm 0.011 $ & (1\%) & $\log({\tt rho})$ \\
~~n=32 & $M/M_{\odot}$                       & $-0.061 \pm 0.025 $ & $(-13\%)$ &  0.01 & $0.004 \pm 0.012 $ & (1\%) &  {\tt mass} \\
\noalign{\smallskip}
asPIC 1.1    & $\log(\rho_{\star}/\rho_{\odot})$ & $-0.078 \pm 0.063 $ & $(-16\%)$ &  0.11 & $0.320 \pm 0.050 $ & (109\%) & $\log({\tt Mass}) - 3\log({\tt Rad})$ \\
~~n=29 & $M/M_{\odot}$                       & $-0.000 \pm 0.038 $ & $(-0\%)$ &  0.49 & $0.167 \pm 0.034 $ & (47\%) &  {\tt Mass} \\
\noalign{\smallskip}
StarHorse 2  & $\log g$ & $-0.031 \pm 0.041 $ & $(-7\%)$ &  0.21 & $0.226 \pm 0.031 $ & (68\%) & {\tt logg50} \\
~~n=34 & $M/M_{\odot}$                       & $-0.085 \pm 0.032 $ & $(-18\%)$ &  0.01 & $0.124 \pm 0.041 $ & (33\%) &  {\tt mass50} \\
\noalign{\smallskip}
Gaia DR3     & $\log(\rho_{\star}/\rho_{\odot})$ & $+0.015 \pm 0.032 $ & $(+4\%)$ &  0.31 & $0.161 \pm 0.025 $ & (45\%) & $\log({\tt mass\_flame\_spec})$\\ 
&&&&&&&$- 3\log({\tt radius\_flame\_spec})$ \\
~~n=28 & $M/M_{\odot}$                       & $-0.036 \pm 0.019 $ & $(-8\%)$ &  0.03 & $0.080 \pm 0.018 $ & (20\%) &  {\tt mass\_flame\_spec} \\
       & $\log g$ & $-0.213 \pm 0.078 $ & $(-39\%)$ &  0.00 & $0.324 \pm 0.075 $ & (111\%) & {\tt logg\_gspspec\_ann} \\
\noalign{\smallskip}
\hline
\end{tabular}
\end{table*}

\begin{figure}
    \centering
    \includegraphics[width=1\linewidth]{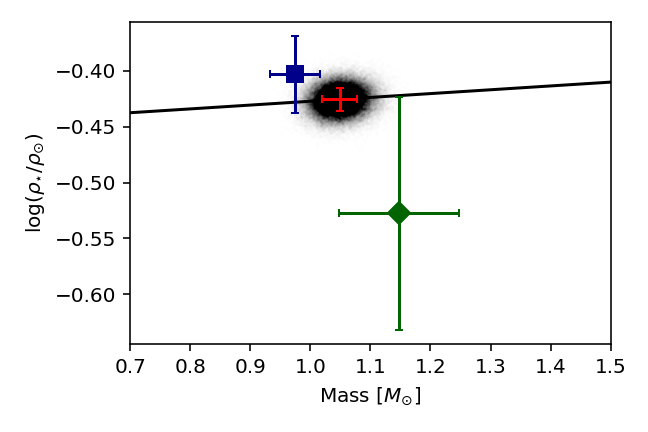}
    \caption{Comparison of the benchmark value of  $\log(\rho_{\star}/\rho_{\odot})$ for EBLM~J1928-38 to values derived from Gaia DR3 (square) and from the asPIC 1.1 catalogue (diamond). The benchmark value is shown as an error bar superposed on values from the Monte Carlo simulation used to obtain the standard error on $\log(\rho_{\star}/\rho_{\odot})$ and the value of    $d\log(\rho_{\star})/dM$. The solid line is centred on the adopted value of the mass and the benchmark value of  $\log(\rho_{\star}/\rho_{\odot})$ and has gradient $d\log(\rho_{\star})/dM_{\star}$.
}
    \label{fig:dlogrhodm_example}
\end{figure}

The largest value of $d\log(\rho_{\star})/dM_{\star}$ in Table~\ref{tab:summary} occurs for EBLM~J1928-38. 
The impact of this weak correlation between $\log(\rho_{\star})$ and the assumed mass is shown in Fig.~\ref{fig:dlogrhodm_example}.
The values of $\log(\rho_{\star})$ from the Gaia~DR3 survey computed from {\tt mass\_flame\_spec} and {\tt radius\_flame\_spec}, and from the asPIC1.1 catalogue are also shown in Fig.~\ref{fig:dlogrhodm_example}.
It can be seen that $d\log(\rho_{\star})/dM_{\star}$ has an almost negligible effect on  tests of estimated $\log(\rho_{\star})$ values for an individual star for any reasonable estimate of the stellar mass. 
Nevertheless, I include the value of  $d\log(\rho_{\star})/dM_{\star}$ in Table~\ref{tab:summary} because it may have a small effect on statistical tests for  catalogues that include high-precision estimates of $\log(\rho_{\star})$ or $\log g$.

\section{Conclusion}
\label{sec:conclusions}

Table~\ref{tab:summary} contains a sample of stars with accurate mass, mean stellar density ($\log \rho_{\star}$) and surface gravity ($\log g$) estimates that can be used to test the accuracy and precision of published estimates of these quantities for FGK-type stars on or near the main sequence.
The mass estimates are based on a re-calibration of a polynomial relation $M_{\star}(\rho_{\star}, {\rm T}_{\rm eff}, {\rm [Fe/H]})$ that provides accurate estimates of the stellar mass with a typical precision better than 5\,per~cent.
The companion stars in the eclipsing binary systems in Table~\ref{tab:summary} are typically more than 8 magnitudes fainter than the primary star in the V band, and so have a negligible impact on the estimates of $\log \rho_{\star}$ and $\log g$ based on data obtained at optical wavelength. 
There may be a very small impact ($\approx 0.5$\,per~cent) on estimates of these quantities that are based on the primary star's luminosity if no correction for the small flux contribution from the faint companion is made. 
The accuracy of the values of $\log \rho_{\star}$ and $\log g$ for eclipsing binaries in Table~\ref{tab:summary} does depend on the accuracy of mass estimates used, but I have shown that these mass estimates are accurate for two test cases where independent estimates of the mass ratio are available, and I have demonstrated how to apply a small correction to these values in cases where an independent estimates of the stellar mass is preferred. 
In most cases, this correction is small enough to have a negligible effect on any statistical tests of stellar parameter estimates compared to the benchmark data in Table~\ref{tab:summary}. 
Several examples of such tests have been given that demonstrate the utility of these benchmark data for detecting small systematic errors in published catalogues of stellar parameters.

\section*{Acknowledgements}

PM thanks the referee for comments that have improved the readability of the manuscript.

PM acknowledges support from UK Science and Technology Facilities Council (STFC) research grant numbers ST/X002047/1 and ST/Y002563/1. 

This work has made use of data from the European Space Agency (ESA) mission
{\it Gaia} (\url{https://www.cosmos.esa.int/gaia}), processed by the {\it Gaia}
Data Processing and Analysis Consortium (DPAC,
\url{https://www.cosmos.esa.int/web/gaia/dpac/consortium}). Funding for the DPAC
has been provided by national institutions, in particular the institutions
participating in the {\it Gaia} Multilateral Agreement.

This paper includes data collected by the {\it Kepler} mission and obtained from the MAST data archive at the Space Telescope Science Institute (STScI). Funding for the {\it Kepler} mission is provided by the NASA Science Mission Directorate. STScI is operated by the Association of Universities for Research in Astronomy, Inc., under NASA contract NAS 5-26555.

This paper includes data obtained through the TESS Guest investigator programs  G06022 (PI Martin), G05024 (PI Martin), G04157 (PI Martin), G03216 (PI Martin) and G022253 (PI Martin).

This paper includes data collected by the TESS mission, which is publicly available from the Mikulski Archive for Space Telescopes (MAST) at the Space Telescope Science Institute (STScI). Funding for the TESS mission is provided by the NASA Explorer Program directorate. STScI is operated by the Association of Universities for Research in Astronomy, Inc., under NASA contract NAS 5-26555.

This research made use of Lightkurve, a Python package for {\it Kepler} and TESS data analysis \citep{2018ascl.soft12013L}.

Based on data obtained from the ESO Science Archive Facility with DOI(s) :  https://doi.org/10.18727/archive/33. 

The author thanks Szilard Csizmadia for making him aware of the early reference to stellar density for stars in eclipsing binary systems by Roberts (1899), and John Southworth for comments on a draft version of the manuscript. 


\section*{Data Availability}
The data underlying this article are available in the following repositories:  
Mikulski Archive for Space Telescopes -- \url{https://archive.stsci.edu/}  (TESS); ESO Science Archive Facility -- \url{https://archive.eso.org/}.



\bibliographystyle{mnras}
\bibliography{allbib} 



\appendix
\section{Data used for re-calibration of empirical mass and radius polynomial functions}
The data used for the re-calibration of the empirical mass and radius polynomial functions described in Section~\ref{sec:recalib} are provided in Table~\ref{tab:debcat}.

\begin{table*}
\centering
\caption{Data used for re-calibration of empirical mass and radius estimates for solar-type stars.}
\label{tab:debcat}
\begin{tabular}{lrrrrrrrr}
\hline\hline
System  & \multicolumn{1}{l}{$P$ [d]} & \multicolumn{1}{l}{[Fe/H]} & 
 \multicolumn{1}{l}{$\log(M_1/M_{\odot})$} &
 \multicolumn{1}{l}{$\log(R_1/R_{\odot})$} &
 \multicolumn{1}{l}{$\log(T_{\rm eff,1}/{\rm K})$} &
 \multicolumn{1}{l}{$\log(M_2/M_{\odot})$} &
 \multicolumn{1}{l}{$\log(R_2/R_{\odot})$} &
 \multicolumn{1}{l}{$\log(T_{\rm eff,2}/{\rm K})$} \\
 \hline
 \noalign{\smallskip}
47 Tuc E32             &  40.913 & $-0.71$ & $-0.0646$  & $ 0.0731$ & 3.7800 & $-0.0826$  & $ 0.0019$ & 3.7750 \\
47 Tuc V69             &  29.540 & $-0.71$ & $-0.0574$  & $ 0.1189$ & 3.7750 & $-0.0661$  & $ 0.0651$ & 3.7770 \\
Kepler-35              &  20.734 & $-0.34$ & $-0.0517$  & $ 0.0122$ & 3.7490 & $-0.0918$  & $-0.1045$ & 3.7160 \\
RW Lac                 &  10.369 & $-0.30$ & $-0.0325$  & $ 0.0741$ & 3.7600 & $-0.0605$  & $-0.0159$ & 3.7450 \\
KIC 6131659            &  17.528 & $-0.23$ & $-0.0289$  & $-0.0533$ & 3.7528 & $-0.1551$  & $-0.1859$ & 3.6794 \\
Kepler-453             &  27.322 & $+0.09$ & $-0.0250$  & $-0.0794$ & 3.7420 &            &           &        \\
TYC 5962-2159-1        &   8.220 & $+0.09$ & $-0.0195$  & $-0.0013$ & 3.7400 & $-0.1713$  & $-0.1612$ & 3.5990 \\
Kepler-47              &   7.448 & $-0.25$ & $-0.0191$  & $-0.0287$ & 3.7510 &            &           &        \\
V565 Lyr               &  18.799 & $+0.28$ & $-0.0020$  & $ 0.0418$ & 3.7480 & $-0.0318$  & $-0.0129$ & 3.7350 \\
V530 Ori               &   6.111 & $-0.12$ & $ 0.0016$  & $-0.0088$ & 3.7770 &            &           &        \\
V818 Tau               &   5.609 & $+0.05$ & $ 0.0105$  & $-0.0348$ & 3.7520 & $-0.1292$  & $-0.1314$ & 3.6330 \\
KIC 7177553            &  17.996 & $-0.05$ & $ 0.0183$  & $-0.0269$ & 3.7630 & $-0.0061$  & $-0.0264$ & 3.7590 \\
Kepler-34              &  27.796 & $-0.07$ & $ 0.0203$  & $ 0.0651$ & 3.7720 & $ 0.0089$  & $ 0.0385$ & 3.7680 \\
TIC 71877648           &  32.394 & $+0.09$ &            &           &        & $ 0.0199$  & $ 0.3462$ & 3.7150 \\
TYC 6296-96-1          &   6.527 & $+0.07$ & $ 0.0327$  & $ 0.0233$ & 3.7730 & $ 0.0278$  & $ 0.0179$ & 3.7690 \\
V963 Cen               &  15.269 & $-0.06$ & $ 0.0339$  & $ 0.1600$ & 3.7640 & $ 0.0315$  & $ 0.1527$ & 3.7650 \\
V568 Lyr               &  14.470 & $+0.28$ & $ 0.0361$  & $ 0.1452$ & 3.7520 & $-0.0822$  & $-0.1072$ & 3.6830 \\
TYC 1243-402-1         &   8.074 & $+0.00$ & $ 0.0374$  & $ 0.0969$ & 3.7750 & $-0.0177$  & $-0.0269$ & 3.7450 \\
QR Hya                 &   5.006 & $-0.01$ & $ 0.0415$  & $ 0.1087$ & 3.7730 & $ 0.0283$  & $ 0.0802$ & 3.7670 \\
AL Dor                 &  14.905 & $-0.10$ & $ 0.0421$  & $ 0.0381$ & 3.7820 & $ 0.0425$  & $ 0.0407$ & 3.7820 \\
V785 Cep               &   6.504 & $-0.06$ & $ 0.0426$  & $ 0.1535$ & 3.7710 & $ 0.0338$  & $ 0.1377$ & 3.7690 \\
EPIC 211409263         &  69.728 & $+0.03$ & $ 0.0457$  & $ 0.0298$ & 3.7850 & $-0.1261$  & $-0.1469$ & 3.6730 \\
TYC 6296-2012-1        &  11.991 & $+0.07$ & $ 0.0496$  & $ 0.0711$ & 3.7880 & $-0.1349$  & $-0.1938$ & 3.6530 \\
KX Cnc                 &  31.220 & $+0.07$ & $ 0.0548$  & $ 0.0223$ & 3.7710 & $ 0.0508$  & $ 0.0250$ & 3.7670 \\
HD 219869              &   6.062 & $-0.39$ & $ 0.0554$  & $ 0.2148$ & 3.7970 & $ 0.0026$  & $ 0.0453$ & 3.7950 \\
TYC 7091-888-1         &  11.658 & $-0.33$ & $ 0.0624$  & $ 0.2634$ & 3.8010 & $-0.1061$  & $-0.1372$ & 3.7320 \\
KIC 7037405            & 207.108 & $-0.27$ &            &           &        & $ 0.0453$  & $ 0.2420$ & 3.7850 \\
EW Ori                 &   6.937 & $+0.05$ & $ 0.0693$  & $ 0.0674$ & 3.7830 & $ 0.0504$  & $ 0.0402$ & 3.7710 \\
KIC 9970396            & 235.299 & $-0.35$ &            &           &        & $ 0.0013$  & $ 0.0449$ & 3.7940 \\
LV Her                 &  18.436 & $+0.08$ & $ 0.0766$  & $ 0.1329$ & 3.7930 & $ 0.0681$  & $ 0.1183$ & 3.7800 \\
AI Phe                 &  24.592 & $-0.14$ & $ 0.0769$  & $ 0.2565$ & 3.7920 & $ 0.0948$  & $ 0.4673$ & 3.7070 \\
LL Aqr                 &  20.178 & $+0.02$ & $ 0.0777$  & $ 0.1209$ & 3.7840 & $ 0.0144$  & $ 0.0009$ & 3.7560 \\
Kepler-1647            &  11.259 & $-0.14$ & $ 0.0866$  & $ 0.2529$ & 3.7930 & $-0.0142$  & $-0.0149$ & 3.7610 \\
KIC 3439031            &   5.952 & $+0.10$ & $ 0.0884$  & $ 0.1484$ & 3.8150 & $ 0.0890$  & $ 0.1471$ & 3.8150 \\
TIC 172900988          &  19.658 & $+0.34$ & $ 0.0926$  & $ 0.1407$ & 3.7820 & $ 0.0799$  & $ 0.1186$ & 3.7770 \\
KIC 8430105            &  63.327 & $-0.46$ &            &           &        & $-0.0897$  & $-0.1253$ & 3.7560 \\
CQ Ind                 &   8.974 & $-0.04$ & $ 0.1036$  & $ 0.1491$ & 3.8100 & $ 0.0528$  & $ 0.0518$ & 3.7880 \\
KIC 7821010            &  24.238 & $+0.10$ & $ 0.1062$  & $ 0.1059$ & 3.8270 & $ 0.0867$  & $ 0.0828$ & 3.8170 \\
V338 Vir               &   5.985 & $-0.10$ & $ 0.1164$  & $ 0.1855$ & 3.8180 &            &           &        \\
CPD -54 810            &  26.132 & $+0.00$ & $ 0.1167$  & $ 0.2853$ & 3.8100 & $ 0.0373$  & $ 0.0724$ & 3.8010 \\
FM Leo                 &   6.729 & $-0.11$ & $ 0.1187$  & $ 0.2107$ & 3.8060 & $ 0.1039$  & $ 0.1785$ & 3.8050 \\
ASAS J085002+1752.5    &   5.226 & $-0.01$ & $ 0.1225$  & $ 0.1962$ & 3.8170 &            &           &        \\
LX Mus                 &  11.751 & $+0.09$ & $ 0.1282$  & $ 0.1287$ & 3.8150 & $ 0.1317$  & $ 0.1449$ & 3.8170 \\
HD 22064               &   9.135 & $-0.05$ & $ 0.1287$  & $ 0.1915$ & 3.8300 &            &           &        \\
RU Cnc                 &  10.173 & $-0.26$ & $ 0.1294$  & $ 0.3400$ & 3.8170 &            &           &        \\
BW Aqr                 &   6.720 & $-0.07$ & $ 0.1377$  & $ 0.2388$ & 3.8100 &            &           &        \\
KIC 9540226            & 175.443 & $-0.31$ &            &           &        & $ 0.0065$  & $ 0.0145$ & 3.7650 \\
BK Peg                 &   5.490 & $-0.12$ &            &           &        & $ 0.0993$  & $ 0.1685$ & 3.8010 \\
HD 149946              &  23.310 & $-0.19$ & $ 0.1526$  & $ 0.3339$ & 3.8220 & $ 0.0573$  & $ 0.0878$ & 3.8190 \\
ASAS J090232-5653.4    &  20.822 & $-0.15$ & $ 0.1572$  & $ 0.1838$ & 3.8470 & $ 0.1374$  & $ 0.1436$ & 3.8250 \\
KIC 8410637            & 408.324 & $+0.02$ &            &           &        & $ 0.1169$  & $ 0.1920$ & 3.8048 \\
KIC 5640750            & 987.398 & $-0.29$ &            &           &        & $ 0.1113$  & $ 0.2679$ & 3.7818 \\
HD 32129               &  16.412 & $+0.19$ & $ 0.1872$  & $ 0.2724$ & 3.8270 & $ 0.0289$  & $-0.0055$ & 3.7600 \\
KIC 11285625           &  10.790 & $-0.58$ & $ 0.1884$  & $ 0.3269$ & 3.8430 & $ 0.0792$  & $ 0.1679$ & 3.8570 \\
KIC 9777062            &  19.230 & $-0.03$ &            &           &        & $ 0.1520$  & $ 0.1886$ & 3.8540 \\
V501 Mon               &   7.021 & $+0.01$ &            &           &        & $ 0.1640$  & $ 0.2019$ & 3.8450 \\
BD +37 410             &  15.535 & $-0.03$ & $ 0.2348$  & $ 0.4622$ & 3.8210 & $ 0.0700$  & $ 0.0741$ & 3.8010 \\
HR 2214                &  23.810 & $-0.23$ & $ 0.2475$  & $ 0.3054$ & 3.8510 &            &           &        \\
TZ For                 &  75.666 & $+0.02$ &            &           &        & $ 0.2918$  & $ 0.5988$ & 3.8030 \\
\hline
\end{tabular}
\end{table*}


\section{Tables of results for transiting exoplanet systems} 
The results of the analysis described in Section~\ref{sec:hj} to estimate the mean stellar density of stars hosting transiting exoplanets are given in Tables \ref{tab:CoRoT-1} to \ref{tab:WASP-43}.
 
\begin{table}
\centering
\caption{Keplerian orbit fit to radial velocity measurements of CoRoT-1.  
 $T_0$ is the Barycentric Dynamical Time (TDB) of mid transit quoted as ${\rm BJD}_{\rm TDB}$.
 $V_0$ is the systemic radial velocity and $K$ is the semi-amplitude of the spectroscopic orbit.
 The root mean square residual (rms) of the residuals from the maximum-likelihood model fit is  $\sigma_{\rm rms}$.
Values preceded by ``='' were not varied in the fit.
\label{tab:CoRoT-1}
}
\begin{tabular}{lcr}
\hline\hline
\multicolumn{1}{@{}l}{Parameter} &  \multicolumn{1}{l}{Units} & \multicolumn{1}{l}{Value}  \\
\hline
\noalign{\smallskip}
\multicolumn{3}{@{}l}{Model parameters} \\
\noalign{\smallskip}
$   T_0               $& [d] & $ =2456268.99119 $ \\
$   P                 $& [d] & $ =1.508968772 $ \\
$ f_c                 $&                  &$   -0.001 \pm   0.021 $ \\
$ f_s                 $&                  &$    -0.01 \pm    0.11 $ \\
$ K                   $& [km\,s$^{-1}$]   &$   0.2311 \pm  0.0089 $ \\
$ V_0                 $& [km\,s$^{-1}$]   &$  23.5046 \pm  0.0059 $ \\
$ \sigma_{\rm jit}    $& [m\,s$^{-1}$]    &$      6.5 \pm     5.8 $ \\
\noalign{\smallskip}
\multicolumn{3}{@{}l}{Data characteristics} \\
\noalign{\smallskip}
$N $ & & 40 \\
$\sigma_{\rm rms} $ & [m\,s$^{-1}$] & 169.7 \\
\noalign{\smallskip}
\multicolumn{3}{@{}l}{Derived parameters} \\
\noalign{\smallskip}
 $ e\sin(\omega)       $&                  &$ -0.0002 \pm   0.0124 $ \\
 $ e\cos(\omega)       $&                  &$  0.0000 \pm   0.0016 $ \\
 $ \psi                $&                  &$    1.001 \pm   0.038 $ \\
\noalign{\smallskip}
\hline
\end{tabular}
\end{table}

\begin{table}
\centering
\caption{Results from the analysis of the Keper light curve of HAT-P-7. $T_0$ is the Julian date of the barycentric dynamical time of mid transit (${\rm BJD}_{\rm TDB}$). The number of observations used in the analysis for each data set are labelled $N_{\it Kepler}$, $N_{\rm rv,1} $, etc. The root mean square residual (rms) of the residuals from the maximum-likelihood model fit for each data set are labelled $\sigma_{\it Kepler} $,
$\sigma_{\rm rv,1} $, etc.  
\label{tab:HAT-P-7}
}
\begin{tabular}{lcr}
\hline\hline
\multicolumn{1}{@{}l}{Parameter} &  \multicolumn{1}{l}{Units} & \multicolumn{1}{l}{Value}  \\
\hline
\noalign{\smallskip}
\multicolumn{3}{@{}l}{Model parameters} \\
\noalign{\smallskip}
$   T_0     $& [d] & $ 2455681.9212037 \pm    0.0000044$ \\
$   P       $& [d] & $ 2.204735442 \pm  0.000000016$ \\
$   D       $&  & $   0.0060150 \pm    0.0000031$ \\
$   W       $&  & $    0.073404 \pm     0.000039$ \\
$   b       $&  & $      0.5003 \pm       0.0010$ \\
$   f_c     $&  & $      -0.008 \pm        0.019$ \\
$   f_s     $&  & $      -0.002 \pm        0.025$ \\
$   q_1     $&  & $       0.255 \pm        0.013$ \\
$   q_2     $&  & $       0.562 \pm        0.011$ \\
$ \sigma_w  $& [ppm] & $  96.57 \pm         0.54$ \\
$   K       $&  [km\,s$^{-1}$] & $      0.2133 \pm       0.0015$ \\
$   V_{0,1}      $& [km\,s$^{-1}$]  & $    -10.5377 \pm       0.0078$ \\
$   V_{0,2}      $& [km\,s$^{-1}$]  & $    -10.5102 \pm       0.0043$ \\
$   V_{0,3}      $& [km\,s$^{-1}$]  & $    -10.5244 \pm       0.0025$ \\
$   V_{0,4}      $& [km\,s$^{-1}$]  & $    -10.4809 \pm       0.0041$ \\
$   V_{0,5}      $& [km\,s$^{-1}$]  & $    -10.5335 \pm       0.0019$ \\
$   \dot{V}   $& [$10^{-5}$ km\,s$^{-1}$\,d$^{-1}$]  & $   5.87 \pm    0.34$ \\
$   \ddot{V}   $& [$10^{-9}$ km\,s$^{-1}$\,d$^{-2}$]  & $8.5 \pm 2.3$ \\
$   \dot{P}    $& [$10^{-10}$ d\,d$^{-1}$] & $    7.16 \pm      0.85$ \\
$   \sigma_{\rm jit}     $& [m\,s$^{-1}$]  & $       9.9 \pm      1.0$ \\
\noalign{\smallskip}
\multicolumn{3}{@{}l}{Data characteristics} \\
\noalign{\smallskip}
$N_{\it Kepler} $ & & 182\,343 \\
$N_{\rm rv,1} $ & & 10 \\
$N_{\rm rv,2} $ & & 8 \\
$N_{\rm rv,3} $ & & 16 \\
$N_{\rm rv,4} $ & & 13 \\
$N_{\rm rv,5} $ & & 42 \\
$\sigma_{\it Kepler} $ & [ppm] & 169.7 \\
$\sigma_{\rm rv,1} $ & [m\,s$^{-1}$] & 5.7 \\
$\sigma_{\rm rv,2} $ & [m\,s$^{-1}$] & 4.1 \\
$\sigma_{\rm rv,3} $ & [m\,s$^{-1}$] & 7.7 \\
$\sigma_{\rm rv,4} $ & [m\,s$^{-1}$] & 16.1 \\
$\sigma_{\rm rv,5} $ & [m\,s$^{-1}$] & 11.1 \\
\noalign{\smallskip}
\multicolumn{3}{@{}l}{Derived parameters} \\
\noalign{\smallskip}
$   R_{\star}/a  $&  & $     0.24162 \pm      0.00020$ \\
$   e       $&  & $     0.00072 \pm      0.00063$ \\
$ \rho_{\star} $& [$\rho_{\odot}$] & $     0.19545 \pm      0.00050$ \\
\noalign{\smallskip}
\hline
\end{tabular}
\end{table}

\begin{table}
\centering
\caption{Keplerian orbit fit to radial velocity measurements of HD~189733. Symbol definitions are the same as those described in the caption to Table~\protect{\ref{tab:HAT-P-7}}.
\label{tab:HD189733}
}
\begin{tabular}{lcr}
\hline\hline
\multicolumn{1}{@{}l}{Parameter} &  \multicolumn{1}{l}{Units} & \multicolumn{1}{l}{Value}  \\
\hline
\noalign{\smallskip}
\multicolumn{3}{@{}l}{Model parameters} \\
\noalign{\smallskip}
$   T_0               $& [d] & $ = 2459446.498519 $ \\
$   P                 $& [d] & $ = 2.218575198$ \\
$f_c              $&                &$   0.0012 \pm   0.0037 $ \\
$f_s              $&                &$   -0.007 \pm    0.043 $ \\
$K                $& [km\,s$^{-1}$] &$   0.2059 \pm   0.0018 $ \\
$V_{0,1}          $& [km\,s$^{-1}$] &$  -2.2638 \pm   0.0036 $ \\
$V_{0,2}          $& [km\,s$^{-1}$] &$  -2.2738 \pm   0.0022 $ \\
$V_{0,3}          $& [km\,s$^{-1}$] &$  -2.2586 \pm   0.0016 $ \\
$\sigma_{\rm jit} $& [m\,s$^{-1}$]  &$     12.2 \pm      1.0 $ \\
\noalign{\smallskip}
\multicolumn{3}{@{}l}{Data characteristics} \\
\noalign{\smallskip}
$N_{\rm rv,1} $ & & 14 \\
$N_{\rm rv,2} $ & & 33 \\
$N_{\rm rv,3} $ & & 52 \\
$\sigma_{\rm rv,1} $ & [m\,s$^{-1}$] & 12.2 \\
$\sigma_{\rm rv,2} $ & [m\,s$^{-1}$] & 9.0 \\
$\sigma_{\rm rv,3} $ & [m\,s$^{-1}$] & 13.7 \\
\noalign{\smallskip}
\multicolumn{3}{@{}l}{Derived parameters} \\
\noalign{\smallskip}
$e\sin(\omega) $&                &$  -0.0001 \pm   0.0021 $ \\
$e\cos(\omega) $&                &$ 0.000036 \pm 0.000087 $ \\
\noalign{\smallskip}
$\psi             $&                &$   1.0002^{+0.0106}_{-0.0023} $ \\
\noalign{\smallskip}
\hline
\end{tabular}
\end{table}

\begin{table}
\centering
\caption{Keplerian orbit fit to radial velocity measurements of HD~209458. 
Symbol definitions are the same as those described in the caption to Table~\protect{\ref{tab:HAT-P-7}}.
\label{tab:HD209458}
}
\begin{tabular}{lcr}
\hline\hline
\multicolumn{1}{@{}l}{Parameter} &  \multicolumn{1}{l}{Units} & \multicolumn{1}{l}{Value}  \\
\hline
\noalign{\smallskip}
\multicolumn{3}{@{}l}{Model parameters} \\
\noalign{\smallskip}
$ T_0               $& [d] & $ = 2455216.405640 $ \\
$ P                 $& [d] & $ = 3.52474859 $ \\
$f_c              $&                &$    0.0021 \pm   0.0079 $ \\
$f_s              $&                &$     0.012 \pm    0.054 $ \\
$K                $& [km\,s$^{-1}$] &$    0.0845 \pm   0.0009 $ \\
$V_{0,1}          $& [km\,s$^{-1}$] &$  -14.7410 \pm   0.0019 $ \\
$V_{0,2}          $& [km\,s$^{-1}$] &$  -14.7404 \pm   0.0010 $ \\
$V_{0,3}          $& [km\,s$^{-1}$] &$  -14.7002 \pm   0.0010 $ \\
$\sigma_{\rm jit} $& [m\,s$^{-1}$]  &$       7.2 \pm      0.9 $ \\
\noalign{\smallskip}
\multicolumn{3}{@{}l}{Data characteristics} \\
\noalign{\smallskip}
$N_{\rm rv,1} $ & & 46 \\
$N_{\rm rv,2} $ & & 141 \\
$N_{\rm rv,3} $ & & 64 \\
$\sigma_{\rm rv,1} $ & [m\,s$^{-1}$] & 13.9 \\
$\sigma_{\rm rv,2} $ & [m\,s$^{-1}$] & 17.5 \\
$\sigma_{\rm rv,3} $ & [m\,s$^{-1}$] & 4.9 \\
\noalign{\smallskip}
\multicolumn{3}{@{}l}{Derived parameters} \\
\noalign{\smallskip}
$e\sin(\omega) $&                &$   0.0002 \pm   0.0035 $ \\
$e\cos(\omega) $&                &$ 0.00008 \pm 0.00027 $ \\
\noalign{\smallskip}
$\psi             $&             &$   0.999^{+0.004}_{-0.017} $ \\
\noalign{\smallskip}
\hline
\end{tabular}
\end{table}

\begin{table}
\centering
\caption{Results from the analysis of the {\it Kepler} light curve of Kepler-1. $T_0$ is the Julian date of the barycentric dynamical time of mid transit (${\rm BJD}_{\rm TDB}$). The number of observations used in the analysis for each data set are labelled $N_{\it Kepler}$, $N_{\rm rv,1} $, etc. The root mean square residual (rms) of the residuals from the maximum-likelihood model fit for each data set are labelled $\sigma_{\it Kepler} $,
$\sigma_{\rm rv,1} $, etc.
\label{tab:Kepler-1}
}
\begin{tabular}{lcr}
\hline\hline
\multicolumn{1}{@{}l}{Parameter} &  \multicolumn{1}{l}{Units} & \multicolumn{1}{l}{Value}  \\
\hline

\noalign{\smallskip}
\multicolumn{3}{@{}l}{Model parameters} \\
\noalign{\smallskip}

$T_0             $ &       [d]          & $ 2455650.0056593 \pm    0.0000021 $\\
$P               $ &       [d]          & $  2.470613369 \pm  0.000000015 $\\
$D               $ &                    & $      0.01627 \pm      0.00011 $\\
$W               $ &                    & $     0.030023 \pm     0.000031 $\\
$b               $ &                    & $       0.8427 \pm       0.0011 $\\
$f_c             $ &                    & $        0.017 \pm        0.030 $\\
$f_s             $ &                    & $        0.002 \pm        0.027 $\\
$q_1             $ &                    & $        0.400 \pm        0.054 $\\
$q_2             $ &                    & $        0.552 \pm        0.019 $\\
$\ell_3          $ &                    & $       0.0175 \pm       0.0057 $\\
$\sigma_w        $ & [ppm]              & $         56.2 \pm          3.3 $\\
$K               $ & [km\,s$^{-1}$]     & $       0.1824 \pm       0.0016 $\\
$V_{0,1}         $ & [km\,s$^{-1}$]     & $      -0.3290 \pm       0.0019 $\\
$V_{0,2}         $ & [km\,s$^{-1}$]     & $      -0.3166 \pm       0.0043 $\\
$V_{0,3}         $ & [km\,s$^{-1}$]     & $       0.1080 \pm       0.0016 $\\
$\sigma_{\rm jit}$ & [m\,s$^{-1}$]      & $          5.2 \pm         1.2 $\\
\noalign{\smallskip}
\multicolumn{3}{@{}l}{Data characteristics} \\
\noalign{\smallskip}
$N_{\it Kepler} $ & & 57\,918 \\
$N_{\rm rv,1} $ & & 10 \\
$N_{\rm rv,2} $ & & 9 \\
$\sigma_{\it Kepler} $ & [ppm] & 239.4 \\
$\sigma_{\rm rv,1} $ & [m\,s$^{-1}$] & 3.9 \\
$\sigma_{\rm rv,2} $ & [m\,s$^{-1}$] & 14.2 \\
\noalign{\smallskip}
\multicolumn{3}{@{}l}{Derived parameters} \\
\noalign{\smallskip}
$R_{\star}/a      $ &                    & $      0.12593 \pm      0.00033 $\\
$e                $ &                    & $       0.0013 \pm       0.0012 $\\
$\rho_{\star}     $ & [$\rho_{\odot}$]   & $       1.0995 \pm       0.0086 $\\
\noalign{\smallskip}
\hline
\end{tabular}
\end{table}

\begin{table}
\centering
\caption{Keplerian orbit fit to radial velocity measurements of WASP-4. 
Symbol definitions are the same as those described in the caption to Table~\protect{\ref{tab:HAT-P-7}}.
\label{tab:WASP-4}
}
\begin{tabular}{lcr}
\hline\hline
\multicolumn{1}{@{}l}{Parameter} &  \multicolumn{1}{l}{Units} & \multicolumn{1}{l}{Value}  \\
\hline
\noalign{\smallskip}
\multicolumn{3}{@{}l}{Model parameters} \\
\noalign{\smallskip}
$ T_0               $& [d] & $ = 2456576.675079 $ \\
$ P                 $& [d] & $ = 1.338231371 $ \\
$f_c              $&                &$    0.001 \pm   0.015 $ \\
$f_s              $&                &$     0.012 \pm    0.062 $ \\
$K                $& [km\,s$^{-1}$] &$    0.2378 \pm   0.0033 $ \\
$V_{0,1}          $& [km\,s$^{-1}$] &$   57.7289 \pm   0.0045 $ \\
$V_{0,2}          $& [km\,s$^{-1}$] &$   57.7452 \pm   0.0023 $ \\
$V_{0,3}          $& [km\,s$^{-1}$] &$   57.7902 \pm   0.0010 $ \\
$\sigma_{\rm jit} $& [m\,s$^{-1}$]  &$       7.3 \pm      1.9 $ \\
\noalign{\smallskip}
\multicolumn{3}{@{}l}{Data characteristics} \\
\noalign{\smallskip}
$N_{\rm rv,1} $ & & 5 \\
$N_{\rm rv,2} $ & & 22 \\
$N_{\rm rv,3} $ & & 17 \\
$\sigma_{\rm rv,1} $ & [m\,s$^{-1}$] &  5.5 \\
$\sigma_{\rm rv,2} $ & [m\,s$^{-1}$] & 20.1 \\
$\sigma_{\rm rv,3} $ & [m\,s$^{-1}$] &  9.2 \\
\noalign{\smallskip}
\multicolumn{3}{@{}l}{Derived parameters} \\
\noalign{\smallskip}
$e\sin(\omega) $&                &$   0.0002 \pm   0.0042 $ \\
$e\cos(\omega) $&                &$ 0.00002 \pm 0.00069 $ \\
\noalign{\smallskip}
$\psi             $&             &$   0.999^{+0.007}_{-0.018} $ \\
\noalign{\smallskip}
\hline
\end{tabular}
\end{table}

\begin{table}
\centering
\caption{Keplerian orbit fit to radial velocity measurements of WASP-43. 
Symbol definitions are the same as those described in the caption to Table~\protect{\ref{tab:HAT-P-7}}.
\label{tab:WASP-43}
}
\begin{tabular}{lcr}
\hline\hline
\multicolumn{1}{@{}l}{Parameter} &  \multicolumn{1}{l}{Units} & \multicolumn{1}{l}{Value}  \\
\hline
\noalign{\smallskip}
\multicolumn{3}{@{}l}{Model parameters} \\
\noalign{\smallskip}
$ T_0               $& [d] & $ = 2455528.868585 $ \\
$ P                 $& [d] & $ = 0.813474079 $ \\
$f_c              $&                &$     0.010 \pm    0.013 $ \\
$f_s              $&                &$     0.025 \pm    0.053 $ \\
$K                $& [km\,s$^{-1}$] &$    0.5438 \pm   0.0048 $ \\
$V_{0,1}          $& [km\,s$^{-1}$] &$   -3.5914 \pm   0.0042 $ \\
$V_{0,2}          $& [km\,s$^{-1}$] &$   -3.5934 \pm   0.0075 $ \\
$V_{0,3}          $& [km\,s$^{-1}$] &$   -3.6212 \pm   0.0043 $ \\
$\sigma_{\rm jit} $& [m\,s$^{-1}$]  &$      18.7 \pm      0.2 $ \\
\noalign{\smallskip}
\multicolumn{3}{@{}l}{Data characteristics} \\
\noalign{\smallskip}
$N_{\rm rv,1} $ & & 26 \\
$N_{\rm rv,2} $ & & 15 \\
$N_{\rm rv,3} $ & & 91 \\
$\sigma_{\rm rv,1} $ & [m\,s$^{-1}$] & 17.1 \\
$\sigma_{\rm rv,2} $ & [m\,s$^{-1}$] & 21.3 \\
$\sigma_{\rm rv,3} $ & [m\,s$^{-1}$] & 38.2 \\
\noalign{\smallskip}
\multicolumn{3}{@{}l}{Derived parameters} \\
\noalign{\smallskip}
$e\sin(\omega) $&                &$ 0.00078 \pm 0.00377 $ \\
$e\cos(\omega) $&                &$ 0.00045 \pm 0.00058 $ \\
\noalign{\smallskip}
$\psi             $&             &$   0.998^{+0.005}_{-0.018} $ \\
\noalign{\smallskip}
\hline
\end{tabular}
\end{table}

\section{KIC 6131659}
\label{sec:kic}
KIC~6131659 is an outlier in the residuals from the empirical polynomial relations computed from the other eclipsing binary stars if the parameters from the version of DEBCat  that was online on 2025-03-05 are used. 
It is apparent from careful inspection of Fig.~10 of \citet{2019AJ....158..106C} that the geometry of the binary system assumed for their analysis of this binary system is not correct and so the radii derived are not reliable. 
The primary eclipse of KIC~6131659 is a transit (annular eclipse) whereas the parameters published by \citeauthor{2019AJ....158..106C} imply incorrectly that it is a partial eclipse.
I have reanalysed the {\it Kepler} short-cadence data for KIC~6131659 using version v43 of the {\sc jktebop}\footnote{\url{http://www.astro.keele.ac.uk/jkt/codes/jktebop.html}} binary star model \citep{2010MNRAS.408.1689S,2023Obs...143...71S}.

I used {\sc lightkurve}\footnote{\url{https://docs.lightkurve.org/}}  \citep{2018ascl.soft12013L} to search the Mikulski Archive for Space Telescopes\footnote{\url{https://archive.stsci.edu/}} (MAST) for {\it Kepler} short-cadence light curves.
The data were downloaded from MAST using {\sc lightkurve} and bad data rejected using the default bit mask. 
Sections of the light curve containing complete eclipses plus some data either side were divided by a straight line fit by least-squares to the data either side of the eclipse, and then exported in a format suitable for analysis using {\sc jktebop}.

The NDE light curve model \citep{1972ApJ...174..617N} on which {\sc jktebop} is based computes eclipses using the approximation that the stars are spherical.  
Since both stars in KIC~6131659 are very nearly spherical and data between the eclipses have not been used, so the ellipsoidal effect was ignored in the analysis of the light curves. 
The reflection effect was also ignored. 
Limb darkening was modelled using the power-2 law and the orbit is assumed to be circular. 
Radial velocity measurements from both \citet{2019AJ....158..106C}  and \citet{2012ApJ...761..157B} are used in the analysis. 
A few data points for each star from \citet{2019AJ....158..106C} observed at phases close to conjunction were ignored because they are clearly affected by systematic errors due to line blending.
The standard errors on each data sets were set equal to the rms of the residuals from the best fit to ensure correct relative weighting of each data set and accurate estimates for the standard errors on the model parameters  estimated using Monte Carlo simulations ({\sc jktebop} task 8).

The parameters of the binary star model are: the sum of the stellar radii in units of
the semi-major axis (fractional radii), $(R_1+R_2)/a$; 
the ratio of the stellar radii, $R_2/R_1$; 
the ratio of the surface brightness at the centre of each stellar disc, $J_0$; 
the orbital inclination, $i$; the time of mid-primary eclipse, $T_0$; 
the orbital period, $P$; 
the radial velocity semi-amplitudes for the spectroscopic orbits of the two stars, $K_1$ and $K_2$; 
third light $\ell_3$; 
systematic velocities of the two stars for the \citeauthor{2019AJ....158..106C} radial velocity data set, $V_{0,1}$ and  $V_{0,2}$;
the offset in the systematic velocities of the two stars in the \citeauthor{2012ApJ...761..157B} relative to the \citeauthor{2019AJ....158..106C} radial velocity data set, $\Delta V_{0,1}$ and  $\Delta V_{0,2}$; 
the $h_1$ limb-darkening parameters for each star, $h_{1,1}$ and $h_{1,2}$;
the zero-point of the normalized {\it Kepler} light curve (not reported here). 
The effect of $h_2$ on the light curve is almost negligible so this parameter was fixed at the nominal value of 0.45 for both stars.

\begin{figure}
    \centering
    \includegraphics[width=1\linewidth]{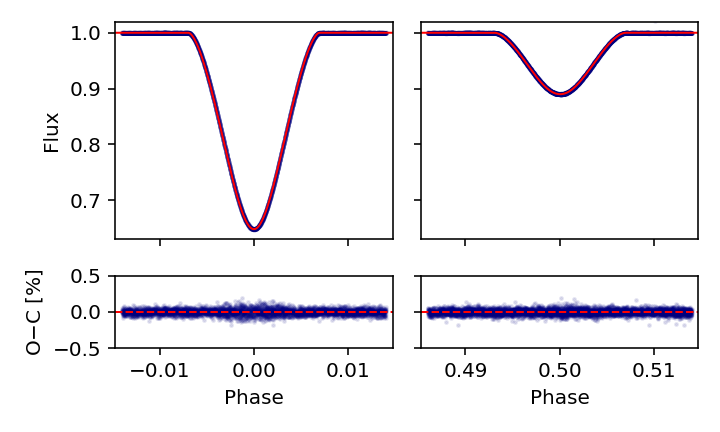}
    \includegraphics[width=1\linewidth]{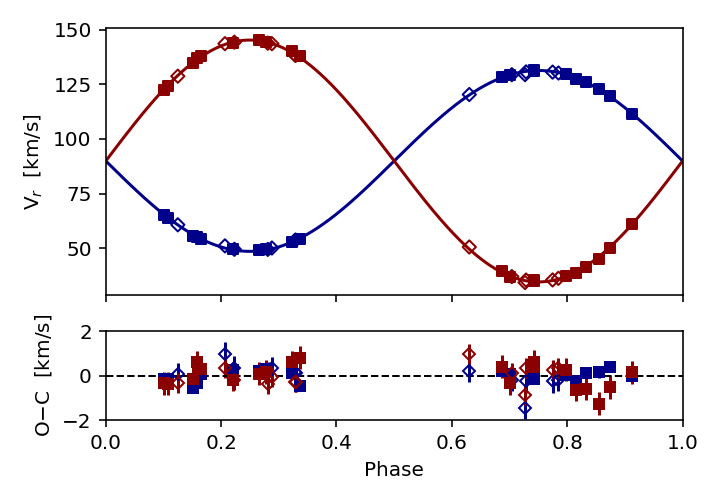}
    \caption{Upper panels: Best-fit {\sc jktebop} model to the {\it Kepler} light curve of KIC~6131659. The residuals from the fit are shown below the main panel.
    Lower panels: Best-fit Keplerian orbits to radial velocity measurements for both stars in KIC~6131659. Radial velocity measurements from \citeauthor{2012ApJ...761..157B} (open diamonds) have been shifted onto a common zero-point with the radial velocity measurements from  \citeauthor{2019AJ....158..106C} (squares). Residuals from the least-squares fit to the radial velocity data are shown below the main panel. }
    \label{fig:kic}
\end{figure}

The best-fit values and standard errors for the model parameters and derived parameters are given in Table~\ref{tab:kic}. 
The best fits to the {\it Kepler} light curve and radial velocities are shown in Fig.~\ref{fig:kic}.
Compared to the values obtained by  \citet{2019AJ....158..106C} the radius of star 1 obtained here is larger by 2.7\,per~cent and the radius of star 2 is smaller by 5.5\,per~cent. These differences are significant compared to the standard errors on these values of 0.14\,per~cent and 0.31\,per~cent, respectively. The quality of the fit shown in  Fig.~\ref{fig:kic} and the results in Table~\ref{tab:kic} show that this re-analysis of the data provides mass and radius estimates for the stars in KIC~6131659 that are both more accurate and more precise.

\begin{table}
\centering
\caption{Results of the analysis of the {\it Kepler} light curve and radial velocity measurements from \citet{2019AJ....158..106C} and \citet{2012ApJ...761..157B} for KIC~6131659 using the {\sc jktebop} binary star model. The rms of the residuals for the radial velocity measurements from \citeauthor{2019AJ....158..106C} and \citeauthor{2012ApJ...761..157B} are reported separately in that order and for each star on the rows labelled $\sigma_{\rm rv,1} $ and $\sigma_{\rm rv,2} $.
The effective temperatures of the two stars are taken from Table~5 of \citeauthor{2012ApJ...761..157B}.
\label{tab:kic}
}
\begin{tabular}{lcr}
\hline\hline
\multicolumn{1}{@{}l}{Parameter} &  \multicolumn{1}{l}{Units} & \multicolumn{1}{l}{Value}  \\
\hline
\noalign{\smallskip}
\multicolumn{3}{@{}l}{Model parameters} \\
\noalign{\smallskip}
$ (R_1+R_2)/a$    &      & $ 0.04592 \pm 0.00004 $ \\
$R_2/R_1$         &      & $ 0.7369 \pm 0.0028 $  \\
$ J_0$            &      & $ 0.3330 \pm 0.0016 $ \\
$ T_0             $& [d] & $  2455748.7944319 \pm 0.0000024 $ \\
$ P               $& [d] & $  17.5278277 \pm 0.0000002 $ \\
$ i               $& [$^{\circ}$]   &$    89.175 \pm   0.004  $ \\
$\ell_3           $&                &$    0.1028 \pm   0.0016 $ \\
$K_1              $& [km\,s$^{-1}$] &$    41.315 \pm   0.060 $ \\
$K_2              $& [km\,s$^{-1}$] &$    55.238 \pm   0.095 $ \\
$V_{0,1}          $& [km\,s$^{-1}$] &$     89.98 \pm    0.06 $ \\
$V_{0,2}          $& [km\,s$^{-1}$] &$     89.96 \pm    0.10 $ \\
$\Delta V_{0,1}   $& [km\,s$^{-1}$] &$    -81.86 \pm    0.16 $ \\
$\Delta V_{0,2}   $& [km\,s$^{-1}$] &$    -81.68 \pm    0.17 $ \\
$h_{1,1}          $&   &$  0.734 \pm   0.003 $ \\
$h_{1,2}          $&   &$  0.668 \pm   0.007 $ \\
\noalign{\smallskip}
\multicolumn{3}{@{}l}{Data characteristics} \\
\noalign{\smallskip}
$N_{\it Kepler} $ & & 9\,897 \\
$N_{\rm rv,1} $ & & 33 \\
$N_{\rm rv,2} $ & & 33 \\
$\sigma_{\it Kepler} $ & [ppm] & 432 \\
$\sigma_{\rm rv,1} $ & [km\,s$^{-1}$] & 0.53, 0.25 \\
$\sigma_{\rm rv,2} $ & [km\,s$^{-1}$] & 0.45, 0.52 \\
\noalign{\smallskip}
\multicolumn{3}{@{}l}{Derived parameters} \\
\noalign{\smallskip}
$M_1           $ & [$M_{\odot}$]      & $ 0.9355 \pm 0.0036 $ \\
$M_2           $ & [$M_{\odot}$]      & $ 0.6997 \pm 0.0023 $ \\
$R_1           $ & [$R_{\odot}$]      & $ 0.8844 \pm 0.0012 $ \\
$R_2           $ & [$R_{\odot}$]      & $ 0.6518 \pm 0.0020 $ \\
$T_{\rm eff,1} $ & [K]                & $   5660 \pm    140 $ \\
$T_{\rm eff,2} $ & [K]                & $   4780 \pm    105 $ \\
$\rho_1        $ & [${\rho_{\odot}}$] & $  1.352 \pm  0.004 $ \\
$\rho_2        $ & [${\rho_{\odot}}$] & $  2.527 \pm  0.023 $ \\
$\log g_1      $ & (cgs)              & $  4.516 \pm  0.001 $ \\
$\log g_2      $ & (cgs)              & $  4.655 \pm  0.003 $ \\
$\log L_1      $ & [$ L_{\odot}$]     & $ -0.141 \pm  0.043 $ \\
$\log L_2      $ & [$ L_{\odot}$]     & $ -0.700 \pm  0.038 $ \\
\hline
\end{tabular}
\end{table}

\bsp	
\label{lastpage}
\end{document}